\documentclass[preprintnumbers,amsmath,amssymb, nofootinbib,tightenlines,preprint]{revtex4}

\usepackage{epsfig,latexsym,cancel,amssymb,amsmath,mathrsfs,verbatim,bbm}
\usepackage{color}
\usepackage{graphicx}

\newcommand{\be}{\begin{equation}}
\newcommand{\ee}{\end{equation}}
\newcommand{\bea}{\begin{eqnarray}}
\newcommand{\eea}{\end{eqnarray}}
\newcommand{\ba}{\begin{array}}
\newcommand{\ea}{\end{array}}
\newcommand{\diracslash}[1]{#1\;\!\!\!\!\!/}
\newcommand{\diracslashb}[1]{#1\;\!\!\!\!/}
\newcommand{\sv}{\left<\sigma v\right>}
\newcommand{\Lint}{\mathscr{L}_{\textrm{int}}}
\newcommand{\mpl}{M_{{pl}}}
\newcommand{\dm}{\delta m}

\long\def\symbolfootnote[#1]#2{\begingroup%
\def\thefootnote{\fnsymbol{footnote}}\footnote[#1]{#2}\endgroup} 

\begin{document}

\title{Oscillating Asymmetric Dark Matter}

\author{Sean Tulin}
\author{Hai-Bo Yu}
\author{Kathryn M. Zurek}
\affiliation{Michigan Center for Theoretical Physics, Department of Physics, University of Michigan, Ann Arbor, MI, USA, 48109}

\date{\today}

\begin{abstract}

We study the dynamics of dark matter (DM) particle-antiparticle oscillations within the context of asymmetric DM.  Oscillations arise due to small DM number-violating Majorana-type mass terms, and can lead to recoupling of annihilation after freeze-out and washout of the DM density.  We derive the density matrix equations for DM oscillations and freeze-out from first principles using nonequilibrium field theory, and our results are qualitatively different than in previous studies.  DM dynamics exhibits particle-vs-antiparticle ``flavor'' effects, depending on the interaction type, analogous to neutrino oscillations in a medium.  ``Flavor-sensitive'' DM interactions include scattering or annihilation through a new vector boson, while ``flavor-blind'' interactions include scattering or $s$-channel annihilation through a new scalar boson, or annihilation to pairs of bosons.  In particular, we find that flavor-sensitive annihilation does not recouple when coherent oscillations begin, and that flavor-blind scattering does not lead to decoherence.

\end{abstract}

\maketitle

\section{Introduction}

The nature and origin of dark matter (DM) remains a fundamental question in our understanding of the Universe. 
Much attention has focused on models where DM is a weakly interacting massive particle (WIMP), with many candidates in theories addressing the gauge hierarchy problem~\cite{Jungman:1995df}.  In these scenarios, the DM density is CP-symmetric and freezes-out when WIMP annihilation falls out of equilibrium (symmetric freeze-out), naturally explaining the observed DM abundance by virtue of the ``WIMP miracle.''  

Alternately, the DM density may be set by its chemical potential \cite{early}, as for early models of technibaryon DM \cite{tbDM}, sterile neutrino DM \cite{nuDM}, and more recently for Asymmetric Dark Matter (ADM) \cite{ADM}, which gives a general prescription for communicating asymmetries between the Standard Model and DM sectors.  This scenario gives rise to a wide range of model building possibilities \cite{General}.  For ADM, the symmetric DM particle-antiparticle density is depleted efficiently through annihilation, and the relic DM abundance is fixed by the initial charge asymmetry (asymmetric freeze-out), associated with a conserved $U(1)_X$ symmetry.  If $U(1)_{X}$ is linked to baryon number, then the observed DM abundance is explained naturally for $\mathcal{O}(5 \, \textrm{GeV})$ DM mass.  
Positive signals from the DAMA/LIBRA~\cite{Bernabei:2010mq}, CoGeNT~\cite{Aalseth:2010vx}, and CRESST~\cite{Angloher:2011uu} experiments also point to this mass range, although it remains unclear whether these results are compatable with each other and with null results from the CDMS-II~\cite{Ahmed:2009zw} and XENON10/100~\cite{Angle:2009xb,Aprile:2010um} experiments~\cite{Arina:2011si}.  Indirect detection signals from dark matter annihilation are typically quenched in ADM models, depending on how efficiently the symmetric DM density is depleted~\cite{Graesser:2011wi}.  However, DM accumulation in stellar systems can provide important constraints on ADM models~\cite{Sandin:2008db}.

The ADM story can change significantly in the presence of tiny $U(1)_{X}$-violating mass terms which give rise to DM particle-antiparticle oscillations. This effect has been discussed within the context of specific models~\cite{Cohen:2009fz,Cai:2009ia,Arina:2011cu,Falkowski:2011xh,Blum:2012nf}, and emphasized more generally in Refs.~\cite{Buckley:2011ye,Cirelli:2011ac}.  If oscillations turn on after freeze-out, the frozen-out DM charge ``thaws'': i.e., the symmetric DM density is repopulated by particle-antiparticle oscillations, and annihilations are reactivated.  Since depletion of the symmetric density during freeze-out requires a large annihilation cross section $\langle \sigma v \rangle \gtrsim 3\times 10^{-26} \, \textrm{cm}^3 \,\textrm{s}^{-1}$, indirect detection signals can be important~\cite{Buckley:2011ye,Cirelli:2011ac}, while stellar ADM bounds can be evaded.  Moreover, residual annihilations can modify the DM relic density substantially~\cite{Buckley:2011ye,Cirelli:2011ac}. On the other hand, if oscillations turn on before freeze-out, the DM charge asymmetry is washed out, and one recovers the usual symmetric freeze-out.  

It is not unreasonable that DM oscillations may occur near the freeze-out epoch.  If $U(1)_X$ is a global symmetry, one expects non-renormalizable $U(1)_X$-violating operators to arise through quantum gravitational effects, suppressed by the Planck scale $\mpl$.  For the case of a fermionic DM state $X$, the lowest dimensional operator is $\phi^\dagger \phi X^2/\mpl$, where $\phi$ is the Higgs field~\cite{Cirelli:2011ac}.  $X$-$\bar{X}$ oscillations turn on when the Majorana mass scale $\langle \phi \rangle^2/\mpl$ is comparable to the Hubble rate $H \sim T^2/\mpl$, corresponding to a temperature $T_{\textrm{osc}} \sim \langle \phi \rangle$ set by the weak scale.

In order to study DM oscillations during freeze-out, one must generalize the usual Boltzmann equations to take into account quantum coherence between particle and antiparticle, provided by the density matrix formalism, as pointed out in Ref.~\cite{Cirelli:2011ac}.  These authors first presented the density matrix equations for oscillating DM, derived by adapting results from neutrino oscillations in the early Universe~\cite{Dolgov:1980cq}.  (The earlier treatment in Ref.~\cite{Buckley:2011ye} treated DM oscillations through a heurtistic modification to the usual Boltzmann equations, without accounting for coherence.)

In this work, we derive these density matrix equations from first principles using nonequilibrium field theory.  We find important qualitative differences with respect to Ref.~\cite{Cirelli:2011ac}: namely, the density matrix structure of the collision term depends crucially the underlying interaction governing DM annihilation.  The situation is analogous to the distinction between flavor-sensitive and flavor-blind interactions in the neutrino context, which determines whether scattering does or does not lead to decoherence by ``measuring'' the flavor of the coherent state.  Here, the two ``flavors'' are particle and antiparticle.  The DM interaction is ``flavor-blind'' or ``flavor-sensitive'' depending on whether the DM bilinear coupling to lighter states is even or odd under charge conjugation ($C$).   One important consequence is, for the flavor-sensitive case, DM annihilation is {\it not} reactivated when oscillations begin, 
contrary to na\"{i}ve expectation.

The remainder of this work is organized as follows.  In Sec.~\ref{sec:setup}, we present the Boltzmann-like density matrix equations describing generic ADM freeze-out and oscillations in the early Universe.  (The details of our derivation, using nonequilibrium field theory, are given in the Appendix.)  In Sec.~\ref{sec:results}, we present our main results.  We discuss the general features of oscillating ADM dynamics, with emphasis on the novel ``flavor'' effects that emerge from our results compared to previous treatments.  We give both numerical and analytical results to illustrate these effects, and we also discuss implications for indirect detection bounds.  Our conclusions are summarized in Sec.~\ref{sec:conc}.

\section{Boltzmann equations}
\label{sec:setup}

\subsection{Basic Setup}

We consider a DM field $X$ that is either a Dirac fermion or complex scalar, with mass $m_X$.  We assume that $X$ carries an approximately conserved charge, corresponding to a $U(1)_X$ symmetry, which is broken through a tiny Majorana-type mass term.  
For fermionic $X$, the Lagrangian is
\be \label{eq:L}
\mathscr{L}_{\textrm{fermion}} = \bar X (i \diracslash{\partial} - m_X) X - \frac{1}{2}\, m_M (\bar X^C X + \bar X X^C) + \mathscr{L}_{\textrm{int}} \; ,
\ee
where $X^C = - i \gamma^2 X^*$ is the charge-conjugated $X$ field.  
For scalar $X$, the Lagrangian is
\be \label{eq:L2}
\mathscr{L}_{\textrm{scalar}} = |\partial_\mu X|^2  - m_X^2 |X|^2 -  \frac{1}{2} \, m_M^2 (X^{C\dagger} X + X^\dagger X^C)  + \mathscr{L}_{\textrm{int}} \; ,
\ee
where $X^C = X^\dagger$.  In Eqs.~\eqref{eq:L} and \eqref{eq:L2}, the interaction $\mathscr{L}_{\textrm{int}}$ describes $X \bar{X}$ or $X X^\dagger$ annihilation into lighter states.

The Majorana-type mass term splits the complex $X$ state into two real states with mass $m_X \pm \delta m$, where 
\be
\delta m \equiv \left\{ \ba{cl} m_M & \quad \textrm{fermionic $X$ case} \\
\frac{m_M^2}{2m_X} & \quad \textrm{scalar $X$ case} \ea \right. \; .
\ee
Since $X$ and $X^C$ are not mass eigenstates for $\delta m \ne 0$, quantum mechanical oscillations occur between $X$ and $X^C$.  To describe the dynamics of ADM freeze-out and oscillations, one must generalize the usual Boltzmann treatment~\cite{Graesser:2011wi} to include quantum coherence between particle and antiparticle states, described by the density matrix
\be
\mathscr{F}_k \sim \left( \ba{cc} \langle a_k^\dagger a_k \rangle & 
\langle b_k^\dagger a_k \rangle \\
\langle a_k^\dagger b_k \rangle &
\langle b_k^\dagger b_k \rangle \ea \right) \; .
\ee
Here, $a_k,a_k^\dagger$ ($b_k,b_k^\dagger$) correspond to (anti)particle creation and annihilation operators for momentum $k$.  
The diagonal elements $\mathscr{F}_{11}$ and $\mathscr{F}_{22}$ correspond to occupation numbers of $X$ and ${X}^C$ states, respectively, while the off-diagonal components govern coherence between them.  

Density matrix equations have been studied previously to describe flavor oscillations in the context of neutrinos~\cite{Dolgov:1980cq} and various baryogenesis scenarios~\cite{Cirigliano:2009yt,Fidler:2011yq,Abada:2006fw,Beneke:2010dz,Garbrecht:2003mn}.
To recast DM particle-antiparticle oscillations in this language, it is helpful to define a DM ``flavor'' doublet $\Psi \equiv (X, X^C)$, where the two ``flavors'' are particle $\Psi_1 \equiv X$ and antiparticle $\Psi_2 \equiv X^C$.  In terms of $\Psi$, the models given in Eqs.~\eqref{eq:L} and \eqref{eq:L2} can be expressed as
\begin{subequations}\label{model2} \begin{align} 
&\mathscr{L}_{\textrm{fermion}} = \frac{1}{2} \bar{\Psi} (i \diracslash{\partial} - M) \Psi + \mathscr{L}_{\textrm{int}} \, , & &M = \left( \ba{cc} m_X & m_M \\ m_M & m_X \ea \right)  \; \\
&\mathscr{L}_{\textrm{scalar}} = \frac{1}{2} |\partial_\mu {\Psi}|^2 - \frac{1}{2}  \Psi^\dagger M^2 \Psi + \mathscr{L}_{\textrm{int}} \, , & &M^2 = \left( \ba{cc} m_X^2 & m_M^2 \\ m_M^2 & m_X^2 \ea \right)  \; .
\end{align}\end{subequations}
The model now appears to be that of two species $\Psi_{1,2}$ that mix via flavor off-diagonal mass terms.  

In the Appendix, we derive the density matrix equation from first principles using nonequilibrium field theory.  For a spatially homogeneous and isotropic expanding Universe, we have (for both fermion and scalar cases)
\be \label{be}
\frac{\partial \mathscr{F}_k}{\partial t} - H k \, \frac{\partial \mathscr{F}_k}{\partial k} = - i \big[ \mathcal{H}_k, \: \mathscr{F}_k \big] + \mathscr{C}_k[\mathscr{F}]  \; ,
\ee
where $H$ is the Hubble rate.  The free Hamiltonian $\mathcal{H}_k$ can be written at leading order in $\delta m$ as
\be \label{ham}
\mathcal{H}_k = \sqrt{k^2 + M^2} = \omega_k\, \mathbbm{1} + \frac{m_X \delta m}{\omega_k} \left( \ba{cc} 0 & 1 \\ 1 & 0 \ea \right)
\ee
with $\omega_k = \sqrt{k^2 + m_X^2}$. (The term proportional to the identity $\mathbbm{1}$ is irrelevant for oscillations.)  

The collision term $\mathscr{C}_k$ depends on the interaction $\mathscr{L}_{\textrm{int}}$.  We assume that $X \bar X$
annihilates into states $f \bar{f}$, where $f$ is a SM or dark sector state.\footnote{We henceforth denote the antiparticle state as $\bar{X}$, inclusive of both fermionic $\bar{X}$ and scalar $X^\dagger$.}  The collision term has two components
\be
\mathscr{C}_k[\mathscr{F}] =  \mathscr{C}_k^a[\mathscr{F}] + \mathscr{C}_k^s[\mathscr{F}] \; ,
\ee
corresponding to annihilation $X(k) \bar{X}(k^\prime) \leftrightarrow f(p) \bar{f}(p^\prime)$
\begin{align}
C_{k}^a[\mathscr{F}]& = - \,\frac{1}{2 \omega_{k}}  \int\! d\Pi_{k^\prime} \label{Cann} 
\int\! d\Pi_{p} \int\! d\Pi_{p^\prime} \: (2\pi)^4 \delta^4 (k+k^\prime-p -p^\prime) \times \frac{1}{2s+1} \sum_{\textrm{spins}}   \\
&  \times\frac{1}{2} \Big( \big\{ \mathscr{F}_k, \, \mathbbm{M}^\dagger_a \bar{\mathscr{F}}_{k^\prime} \mathbbm{M}_a \big\}\, (1\pm f_p) (1 \pm \bar f_{p^\prime})
- \big\{ \mathbbm{1} \pm  \mathscr{F}_k, \, \mathbbm{M}_a^\dagger (\mathbbm{1}\pm \bar{\mathscr{F}}_{k^\prime}) \mathbbm{M}_a \big\} f_p \bar f_{p^\prime} \Big) \notag \; ,
\end{align}
and scattering $X(k) f(p) \leftrightarrow X(k^\prime) f(p^\prime)$ and $X(k) \bar f(p) \leftrightarrow X(k^\prime) \bar f(p^\prime)$
\begin{align} 
C_k^s[\mathscr{F}] &= - \,\frac{1}{2 \omega_{k}}   \int\! d\Pi_{k^\prime} \label{Cscat}
\int\! d\Pi_{p} \int\! d\Pi_{p^\prime} \: (2\pi)^4 \delta^4 (k+p-k^\prime -p^\prime) \times \frac{1}{2s+1} \sum_{\textrm{spins}} \\
&  \times   \frac{1}{2} \Big( \big\{ \mathscr{F}_k, \, \mathbbm{M}^\dagger_s (\mathbbm{1}\pm \mathscr{F}_{k^\prime}) \mathbbm{M}_s \big\}\, f_p (1 \pm f_{p^\prime})
- \big\{ \mathbbm{1}\pm \mathscr{F}_k, \, \mathbbm{M}_s^\dagger \mathscr{F}_{k^\prime} \mathbbm{M}_s \big\} (1\pm f_p) f_{p^\prime} \Big) \notag \\
& + \; \big( \, f \to \bar{f}\, \big) \notag \; ,
\end{align}
with phase space measure $d\Pi_k \equiv d^3k /\left((2\pi)^3 \,2\omega_k\right)$ and $+$ ($-$) sign for bosons (fermions), and where $f_p$ ($\bar f_p$) is the $f$ (anti)particle distribution function, with momentum $p$.  Our expression for $\mathscr{C}_k$ averages over the spin $s$ of $X(k)$, given by the $(2s+1)^{-1}$ factor, and sums over spins for all other states.  The annihilation term involves the ``barred'' density matrix
\be \label{barredF}
\bar{\mathscr{F}} \equiv \left( \ba{cc} \mathscr{F}_{22} & \mathscr{F}_{12} \\ \mathscr{F}_{21} & \mathscr{F}_{11} \ea \right) \; ,
\ee
the form for which is derived in the Appendix.  

In the density matrix equations, the annihilation and scattering amplitudes become matrices in flavor-space, given by
\begin{subequations} \begin{align}
\mathbbm{M}_a &= \left( \ba{cc} \mathcal{M}(X\bar{X}\leftrightarrow f\bar{f}) & 0 \\ 0 &  \mathcal{M}(X^C\bar{X}^C\leftrightarrow f\bar{f}) \ea \right)  \\ \mathbbm{M}_s &= \left( \ba{cc} \mathcal{M}(X f\leftrightarrow Xf) & 0 \\ 0 &  \mathcal{M}(X^C f\leftrightarrow X^C f) \ea \right) \; ,
\end{align}\end{subequations}
respectively, where $\mathcal{M}$ is the usual matrix element.  If $\mathbbm{M}_{a,s}$ is proportional to the identity, these interactions are ``flavor-blind''; otherwise interactions are ``flavor-sensitive.''  The distinction between flavor-blind and flavor-sensitive turns out to be critically important for oscillating DM.

In the absence of coherence ($\mathscr{F}_{12} = \mathscr{F}_{21} = 0$), it is straightforward to see that Eqs.~\eqref{Cann} and \eqref{Cscat} reproduce the usual Boltzmann collision terms.  $\mathscr{F}_{11}$ ($\mathscr{F}_{22}$) corresponds to the $X$ ($\bar{X}$) occupation number, and $\mathbbm{M}_{a,s}$ and $\mathbbm{M}_{a,s}^\dagger$ factor out, giving the usual squared matrix elements $\sum |\mathcal{M}|^2$.  

To consider a concrete example, we take $X$ and $f$ to be fermions, coupled through an effective contact interaction
\be 
\Lint = \frac{G_X}{\sqrt{2}}\, \bar{X} \Gamma^a X \, \bar{f} \Gamma_a f \; , \label{contact}
\ee
with coupling $G_X$, obtained by integrating out a heavy mediator.  The Dirac structure is given by $\Gamma^a$: scalar $\Gamma^S = 1$, pseudoscalar $\Gamma^A = \gamma_5$, vector $\Gamma^V = \gamma^\mu$, axial vector $\Gamma^A = \gamma^\mu \gamma_5$, and tensor $\Gamma^T = \sigma^{\mu\nu} \equiv \frac{i}{2} [ \gamma^\mu, \gamma^\nu]$.  In terms of $\Psi$, Eq.~\eqref{contact} becomes
\be
\Lint = \frac{G_X}{2 \sqrt{2}}\, \bar{\Psi} \Gamma^a O_\pm \Psi \, \bar{f} \Gamma_a f \; , \label{contact2}
 \quad O_\pm \equiv \left( \ba{cc} 1 & 0 \\ 0 & \pm 1 \ea \right) \; ,
\ee
where the $\pm$ in $O_\pm$ corresponds to the transformation property of $\Lint$ under $X \to X^C$.  Scalar, pseudoscalar, and axial-vector interactions are flavor-blind ($+$), while vector and tensor interactions are flavor-sensitive ($-$).  The amplitude matrices factorize as
\be
\mathbbm{M}_a = \mathcal{M}(X \bar{X} \to f \bar f) \, O_\pm \, , \quad
\mathbbm{M}_s = \mathcal{M}(X f \to X f) \, O_\pm \; . 
\ee
In a more general case with mixed $C$ ({\it e.g.}, $\Gamma^a = g_V \gamma^\mu + g_A \gamma^\mu \gamma_5$), both $O_\pm$ contribute:
\begin{subequations} \label{factor}
\begin{align}
\mathbbm{M}_a &= \mathcal{M}_+(X \bar{X} \to f \bar f) \, O_+ + \mathcal{M}_-(X \bar{X} \to f \bar f) \, O_- \\
\mathbbm{M}_s &= \mathcal{M}_+(X f \to X  f) \, O_+ + \mathcal{M}_-(X f \to X  f) \, O_- 
\end{align}
\end{subequations}
where $\mathcal{M}_+$ ($\mathcal{M}_-$) is the part of the matrix element proportional to $g_{A}$ ($g_{V}$).\footnote{To be clear, we emphasize that $C = \pm$ does {\it not} refer to the $C$-transformation of $\Lint$ in the usual sense, where one transforms all fields entering $\Lint$ under $C$.  Here, $C = \pm$ refers to the parity of $\Lint$ under $X \to X^C$, while keeping the other fields untransformed.  In this latter sense, we identify $C$-even (odd) interactions as corresponding to flavor-blind (sensitive) collisions.}  

Eq.~\eqref{factor} corresponds to the most general form for the amplitude matrices for any interaction $\Lint$.  Although our results were derived for a contact interaction (see Appendix), it is straightforward to adapt our results to any $\Lint$ by using the appropriate matrix elements $\mathcal{M}_{a,s}$.  The sign of $O_\pm$ is determined by $\mathcal{M}_a \to \pm \mathcal{M}_a$ under $X \to X^C$.  One important example is $X \bar{X}$ annihilation to light dark sector bosons (which then decay to SM states); this case has $O_+$.

\subsection{Nonrelativistic limit}

The density matrix equation can be simplified considerably if $X,\bar{X}$ are nonrelativistic, as expected during and after freeze-out.  
The usual prescription in the single flavor case is to integrate the Boltzmann equation and to express everything terms of total number densities.  Analogously, we define a ``number density matrix''
\be
n \equiv (2s+1) \int\!\! \frac{d^3 k}{(2\pi)^3} \: \mathscr{F}_k = \left( \ba{cc} n_{11} & n_{12} \\ n_{21} & n_{22} \ea \right) \,  , \quad 
\bar{n} \equiv (2s+1) \int\!\! \frac{d^3 k}{(2\pi)^3} \: \bar{\mathscr{F}}_k = \left( \ba{cc} n_{22} & n_{12} \\ n_{21} & n_{11} \ea \right)  \, ,
\ee
where the $(2s+1)$ factor accounts for spin.  To evaluate the integrated collision term $\int {d^3 k}/(2\pi)^3 \,  \mathscr{C}_k$ in terms of $n$ (and $\bar{n}$), we take as an ansatz 
\be
\mathscr{F}_k = e^{-\omega_k/T} \, \frac{ n}{n_{\textrm{eq}}} \; , \quad \bar{\mathscr{F}}_k = e^{-\omega_k/T} \, \frac{\bar n}{n_{\textrm{eq}}} \; ,
\ee
assuming that the momentum dependence of $(\mathscr{F}_k)_{ij}$ can be characterized by a Maxwell-Boltzmann factor independent of $ij$, where $n_{\textrm{eq}} \equiv (2s+1) \int d^3 k/(2\pi)^3 \, \exp(-\omega_k/T)$.  We expect this ansatz to be valid since the free Hamiltonian, in the nonrelativistic limit, becomes
\be
\mathcal{H}_0 = \left( \ba{cc} m_X & \delta m \\ \delta m & m_X \ea \right) \; ,
\ee
giving an oscillation frequency $\omega_{\textrm{osc}} = 2 \delta m$ that is approximately independent of $k$, modulo $\mathcal{O}(k^2/m_X^2)$ corrections.  Moreover, we take a general structure for the amplitude matrices, given in Eq.~\eqref{factor}.

Taking the integral $(2s+1) \int d^3 k/(2\pi)^3$ of Eq.~\eqref{be}, the integrated density matrix equation is
\be \label{be2}
\frac{\partial n}{\partial t} + 3 H n = - i \big[ \mathcal{H}_0, \, n \big] 
- \frac{\Gamma_{\pm}}{2}\,  \big[ O_\pm, \big[ O_\pm,\, n \big] \big]
 -  \langle \sigma v \rangle_\pm  \Big( \frac{1}{2} \big\{ n, \, O_\pm \,\bar{n} \, O_\pm \big\} -  n_{\textrm{eq}}^2 \Big) \; .
\ee
The terms on the right-hand side correspond to oscillations, scattering, and annihilation, respectively. 
The $\pm$ denotes collision terms from flavor-blind ($+$) and flavor-sensitive ($-$) interactions, and in general both types contribute.  The usual thermally-averaged cross section is $\langle \sigma v \rangle \equiv \langle \sigma v \rangle_+ + \langle \sigma v \rangle_-$, but the separate $C$-even ($\langle \sigma v \rangle_+$) and $C$-odd ($\langle \sigma v \rangle_-$) contributions
have a different matrix structure in the anticommutator term, due to $O_\pm$.  There are no $O_+ O_-$ cross terms: the different $C$ amplitudes do not interfere, since a particle-antiparticle wavefunction is an eigenstate of $C$.  
The total thermally-averaged elastic scattering rate for $X f \leftrightarrow X f$ plus $ X \bar f \leftrightarrow X \bar f$ is
\begin{align}
\Gamma_\pm = 2 \, \frac{1}{n_{\textrm{eq}}} &\int \! d\Pi_k \int \! d\Pi_p \int \! d\Pi_{k^\prime} \int \! d\Pi_{p^\prime}
\, (2\pi)^4 \delta^4(k+p-k^\prime-p^\prime) \notag \\
& \quad \times \sum_{\textrm{spins}} | \mathcal{M}_\pm(X f \leftrightarrow X f)|^2 \, e^{-\omega_k/T} f_p (1\pm f_{p^\prime}) \;
\label{scatrate} .
\end{align}
No $O_+ O_-$ cross terms arise for $f_p = \bar{f}_p$, which we have assumed in Eq.~\eqref{scatrate}.  Moreover, since $O_+ = \mathbbm{1}$ commutes with any $n$, only flavor-sensitive scattering contributes to Eq.~\eqref{be2}.

Next, we define the comoving number density matrix ${Y} \equiv n/s$ (and $\bar{Y} \equiv \bar{n}/s$)~\cite{Cirelli:2011ac}, where $s = 2\pi^2/45 \, g_{*S}(T) \, T^3$ is the entropy density\footnote{Below, we use $s$ to denote entropy density, not to be confused with particle spin $s$.} and $g_{*S}$ counts the effective number of relativistic degrees of freedom.  We take the notation $'\equiv\left( 1 -  \frac{1}{3} \frac{\partial \ln g_{*S}}{\partial \ln x} \right)^{-1} \frac{d}{dx}$, and rewrite Eq.~\eqref{be2} as
\be \label{be3}
Y'(x)= - \frac{i}{Hx} \big[ \mathcal{H}_0, \, {Y} \big] - \frac{\Gamma_\pm}{2 Hx}\, \big[ O_\pm, \big[ O_\pm,\, {Y} \big] \big]
 -  \frac{s \langle \sigma v \rangle_\pm}{Hx}  \, \Big( \frac{1}{2} \big\{ {Y}, \, O_\pm \,\bar{Y} \, O_\pm \big\} -  {Y}_{\textrm{eq}}^2 \Big) 
\ee
where $x \equiv m_X/T$ and ${Y}_{\textrm{eq}} \equiv n_{\textrm{eq}}/s$.  We also denote the $X, \bar X$ comoving number densities as $Y_X \equiv Y_{11}$ and $Y_{\bar X} \equiv Y_{22}$.  

Eq.~\eqref{be3} is the master Boltzmann equation for oscillating DM.  Similar results were presented in Ref.~\cite{Cirelli:2011ac}, but do not capture the correct matrix structure of the annihilation and scattering terms, nor the distinction between flavor-blind and flavor-sensitive interactions.  These subtleties are qualitatively important in studying oscillating DM.

\section{Discussion and Results}
\label{sec:results}

For oscillating DM, freeze-out dynamics and indirect detection signals can depend crucially on whether the interactions responsible for DM annihilation and elastic scattering are flavor-sensitive or flavor-blind.  We now discuss these issues in detail.  We first consider the annihilation and scattering terms, and then we present numerical and analytical solutions to the density matrix equations which illustrate our discussion.  Lastly, we briefly mention implications for indirect detection signals.

\subsection{Annihilation}

First, we consider the annihilation term; expanding the anticommutator, we have\footnote{The annihilation term given in Ref.~\cite{Cirelli:2011ac} is different in two respects: the authors (i) set $O_\pm = \mathbbm{1}$ for all types of interactions, and (ii) use a different form for $\bar{Y}$ where $\bar{Y}_{12,21} = - Y_{21,12}$.}
\begin{subequations}
\begin{align}
&\textrm{flavor-blind:}  & \frac{1}{2} \big\{ Y, \, O_+ \bar{Y} O_+ \big\} \;=& \;\left( \ba{cc} Y_{11} Y_{22} + Y_{12} Y_{21} & Y_{11} Y_{12} + Y_{12} Y_{22} \\  Y_{21} Y_{11} + Y_{22} Y_{21} & Y_{11} Y_{22} + Y_{12} Y_{21} \ea \right)\\
&\textrm{flavor-sensitive:}  & \frac{1}{2} \big\{ Y, \, O_- \bar{Y} O_- \big\} \;=&\; \left( \ba{cc} Y_{11} Y_{22} - Y_{12} Y_{21} & 0 \\  0 & Y_{11} Y_{22} - Y_{12} Y_{21} \ea \right) \; .
\end{align}
\end{subequations}
The two types of interactions couple very differently to $Y_{ij}$.  However, in the absence of coherence ($Y_{12}, Y_{21} \to 0$), both interactions give the same (usual) result proportional to $Y_X Y_{\bar X}$.  The distinction between flavor-blind and flavor-sensitive is only relevant in the presence of coherence.

If oscillations turn on after freeze-out, one na\"{i}vely expects annihilation to be reactivated as $X$ oscillates into $X^C$, repopulating $X^C$.  This expectation turns out to be false for flavor-sensitive annihilation.  In this case, annihilation only couples to $Y$ through $\textrm{det}(Y) = Y_{11} Y_{22} - Y_{12} Y_{21}$.  Because $\det( [\mathcal{H}_0,Y]) = 0$, oscillations do not ``source'' flavor-sensitive annihilation.  As long as DM is coherently oscillating, annihilation is not reactivated.

\begin{figure}[t]
\begin{center}
\includegraphics[scale=1]{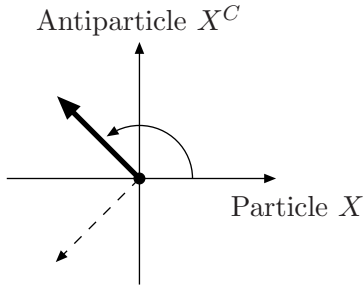}
\caption{\it DM freezes out as a pure $X$ state, and then precesses in $X$-$X^C$ space due to coherent DM oscillations.  
For nonrelativistic DM, all states precess approximately uniformly, shown by the solid arrow.  
For flavor-sensitive interactions, a state only annihilates with an orthogonal one, shown by the dashed arrow, which is not populated.\label{precess}}
\end{center}
\end{figure}

This result stems from a simple symmetry argument.  Annihilation occurs through a two-particle state characterized by spin, spatial, and flavor (i.e., $X,X^C$) wavefunctions.  Moreover, since both $X$ and $X^C$ must be present to annihilate, and particle-antiparticle wavefunctions are eigenstates of $C$, the total wavefunction has eigenvalue $C=(-1)^{L+S}$, where $L$ is the total angular momentum, and $S$ is the total spin.  Boson (fermion) statistics requires that the total wavefunction be (anti)symmetric. For all choices of $L$ and $S$, this implies that $C$-even (odd) interactions have (anti)symmetric flavor wavefunctions, according to the following table.
\begin{table}[h!]
\begin{center}
\begin{tabular}{c|c|ccc|c}
 & $\;C\;$ & $\;S\;$ & $\;\;\;L\;\;\;$ & \; flavor \; & \; total \; \\
\hline
scalar $X$ & $+$ & --- & even & even & even \\
& $-$ & --- & odd & odd & even \\
\hline
fermion $X$ & $+$ & \; 0 (odd) & even & even & odd\\
& $-$ & \; 0 (odd) & odd & odd & odd\\
 & $+$ & \; 1 (even) & odd & even & odd\\
& $-$ & \; 1 (even) & even & odd & odd\\
\hline
\end{tabular}
\end{center}
\end{table}
\newline If oscillations turn on when DM is nonrelativistic, all states precess uniformly (with $\omega_{\textrm{osc}} \approx 2\delta m$) and only one pure state is populated, illustrated in Fig.~\ref{precess}.  Therefore, only a symmetric flavor wavefunction can be nonvanishing.  Flavor-sensitive annihilation, requiring an antisymmetric flavor wavefunction, remains frozen-out.  Once the coherence is broken, DM is no longer a pure state, and annihilation commences.

Even in the absence of collisions, decoherence can occur within the thermal DM ensemble.
Since DM particles have a thermal distribution in momentum $k$, different momentum modes can go out of phase, due to $k$-dependent corrections to the oscillation frequency, given by
\be
\omega_{\textrm{osc}}(k) = 2\delta m \left( 1  -  \frac{k^2}{2 m_X^2} + \mathcal{O}\left(\frac{k^4}{m_X^4} \right) \right) \; .
\ee
This thermal effect breaks the coherence of the ensemble and leads to annihilation.  A rigorous treatment of this effect requires, however,  solving Eq.~\eqref{be} for $\mathscr{F}_k$ directly, which is beyond the scope of this work.  The integrated density matrix equation, given in Eq.~\eqref{be3}, neglects $\mathcal{O}(k^2/m_X^2)$ corrections to $\omega_{\textrm{osc}}$ and does not include this effect.

To estimate the time scale when flavor-sensitive annihilation begins, we consider a DM state $X_k$ with momentum $k$.  At time $t=0$, we have $X_k(0) = | X \rangle$, and the state evolves according to $X_k(t)=\cos(\omega_{\textrm{osc}} t/2)|X\rangle-i\sin(\omega_{\textrm{osc}} t/2)|X^C\rangle$, neglecting an overall phase $\exp(-i \omega_k t)$.  An antisymmetric flavor wavefunction can be composed from two states with momentum $k,k^\prime$ as follows:
\begin{eqnarray}
X_k(t)\otimes X_{k^\prime}(t)-X_{k^\prime}(t)\otimes X_k(t)= i \sin( \Delta \omega_{\textrm{osc}}t/2) \big(|X\rangle\otimes|X^c\rangle-|X^c\rangle\otimes|X\rangle \big) \,,
\end{eqnarray}
where $\Delta \omega_{\textrm{osc}} \equiv \omega_{\textrm{osc}}(k) - \omega_{\textrm{osc}}(k^\prime)$.  The wavefunction becomes nonvanishing and annihilation commences for $t \gtrsim \tau_{\textrm{dec}}$, with decoherence time scale $\tau_{\textrm{dec}} \equiv |\Delta \omega_{\textrm{osc}}^{-1}|  \sim {1}/{(\delta m v^2)}$, where $v$ is the typical DM velocity.  Since $v \ll 1$ for nonrelativistic DM, the onset of flavor-sensitive annihilation can be significantly delayed compared to when oscillations begin.

\subsection{Elastic scattering}

Next, we consider DM elastic scattering with the thermal plasma.  For a flavor-sensitive interaction, the scattering term is
\be
\frac{\Gamma_-}{2} \big[ O_-, \big[ O_-,\, {Y} \big] \big] = 2\Gamma_- \left( \ba{cc} 0 & Y_{12} \\ Y_{21} & 0 \ea \right) \; ,
\label{eq:scattering}
\ee
which damps $Y_{12}, Y_{21} \to 0$ and causes decoherence of DM oscillations. 
Typically, oscillations begin when $\omega_{\textrm{osc}} \sim H$.  However, if $\Gamma_- > H$, then scattering plays an important dynamical role.  At first, when $\Gamma_- > \omega_{\textrm{osc}} > H$, coherent oscilations do not occur due to the quantum Zeno effect.
The $X$ asymmetry does not oscillate into $X^C$ because flavor-sensitive interactions rapidly ``measure'' the state to be $X$ before $X \to {X}^C$ can occur.  ADM does not thaw, and annihilation remains frozen-out.
Next, when $\Gamma_- \sim \omega_{\textrm{osc}}$, the quantum Zeno effect no longer occurs.  The pure $X$ state, through oscillations and decoherence from scattering, is reduced to a fully mixed $X$-$X^C$ system ($\mathscr{F}_k \propto \mathbbm{1}$).  Annihilation commences for $\omega_{\textrm{osc}} > \Gamma_-$.

In the case of a flavor-blind interaction, scattering does not lead to decoherence.  Since $O_+ = \mathbbm{1}$, the scattering term vanishes.\footnote{In contrast, Ref.~\cite{Cirelli:2011ac} adopts a scattering term as in Eq.~\eqref{eq:scattering} for all types of interactions.}  This result is known from neutrino physics: purely flavor-blind (i.e. neutral current) iso-momentum $\nu$ scattering on nonrelativistic targets does not lead to decoherence.\footnote{For the neutrino case, this effect is not preserved when one includes charge-current forward scattering effects; see discussion in Ref.~\cite{Raffelt:1992bs} Appendix A.}  Here, scattering does not measure the state, leaving the wavefunction uncollapsed; coherence is preserved.

For the contact interaction given in Eq.~\eqref{contact}, the scattering rate is correlated with the annihilation cross section.  For example, for a vector interaction $\Gamma^a = \gamma^\mu$ we have
\be
\sv_- = \frac{g_f G_X^2 m_X^2}{2\pi} \,, \quad \Gamma_- = \frac{ 7 g_f G_X^2 T^5 \pi}{60} \; ,
\ee
for $m_f = 0$ and where $g_f$ counts the $f$ degrees of freedom ({\it e.g.} color).  Within a more general theory of DM, the scattering and annihilation rates are less correlated.  The scattering rate can be suppressed compared to annihilation if the latter is resonantly enhanced or has final states that are heavy $(m_X > m_f \gtrsim m_X/20)$ such that scattering is Boltzmann suppressed during freeze-out; or, scattering can be compartively enhanced if there exists a dark sector thermal bath with many light states to scatter from.  

\subsection{Numerical results}

From a model-building perspective, ADM oscillations offer an appealing mechanism to allow DM masses at the weak scale, well above the natural ADM mass scale of 5 GeV.  If the oscillation parameter satisfies $\delta m \sim 10^{-10}\, {\rm eV} \times (m_X/10 \, {\rm GeV})^2$, then oscillations can begin during the freeze-out epoch, potentially allowing for residual annihilation to deplete the DM density below its asymmetric abundance.  Here, we present numerical solutions to the density matrix equations in order to illustrate these dynamics, focusing on the difference between flavor-blind versus flavor-sensitive interactions.  

\begin{figure}[t]
\begin{center}
\includegraphics[scale=0.6]{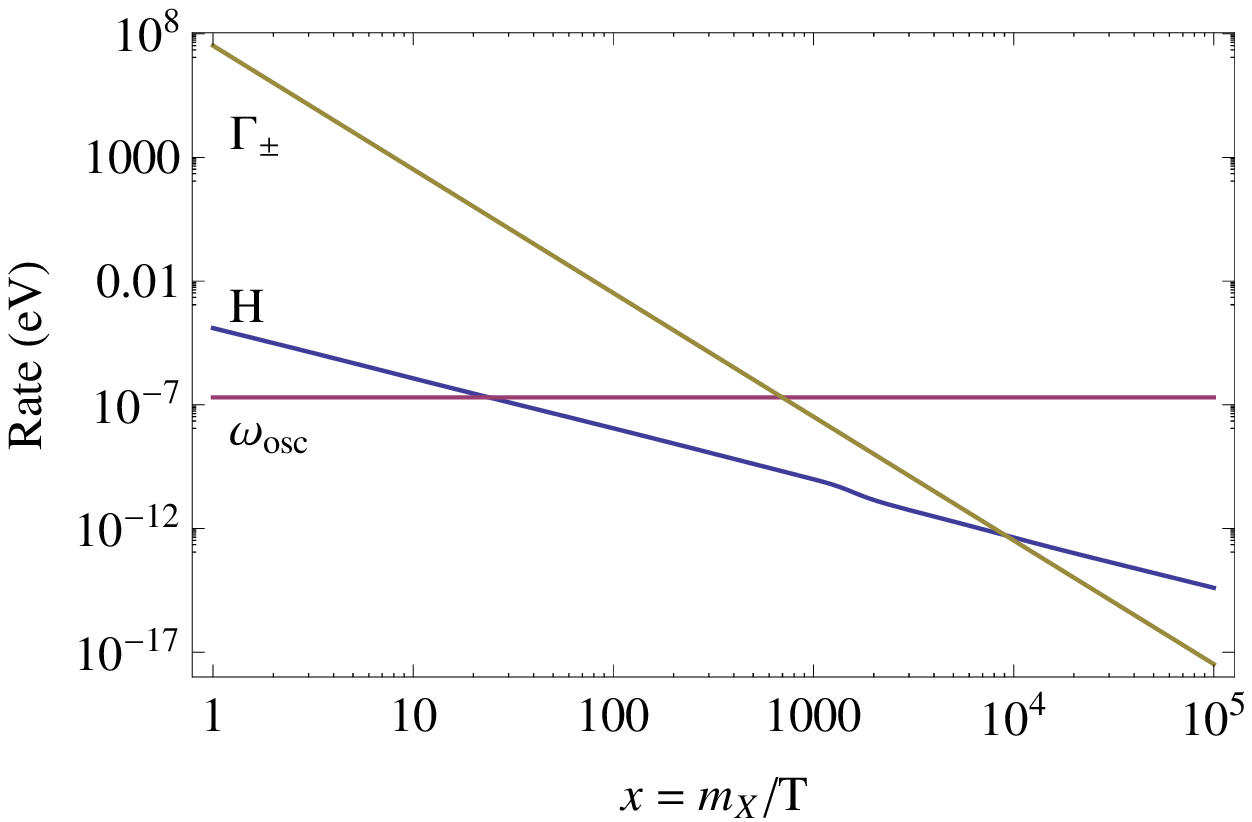} ~ \includegraphics[scale=0.6]{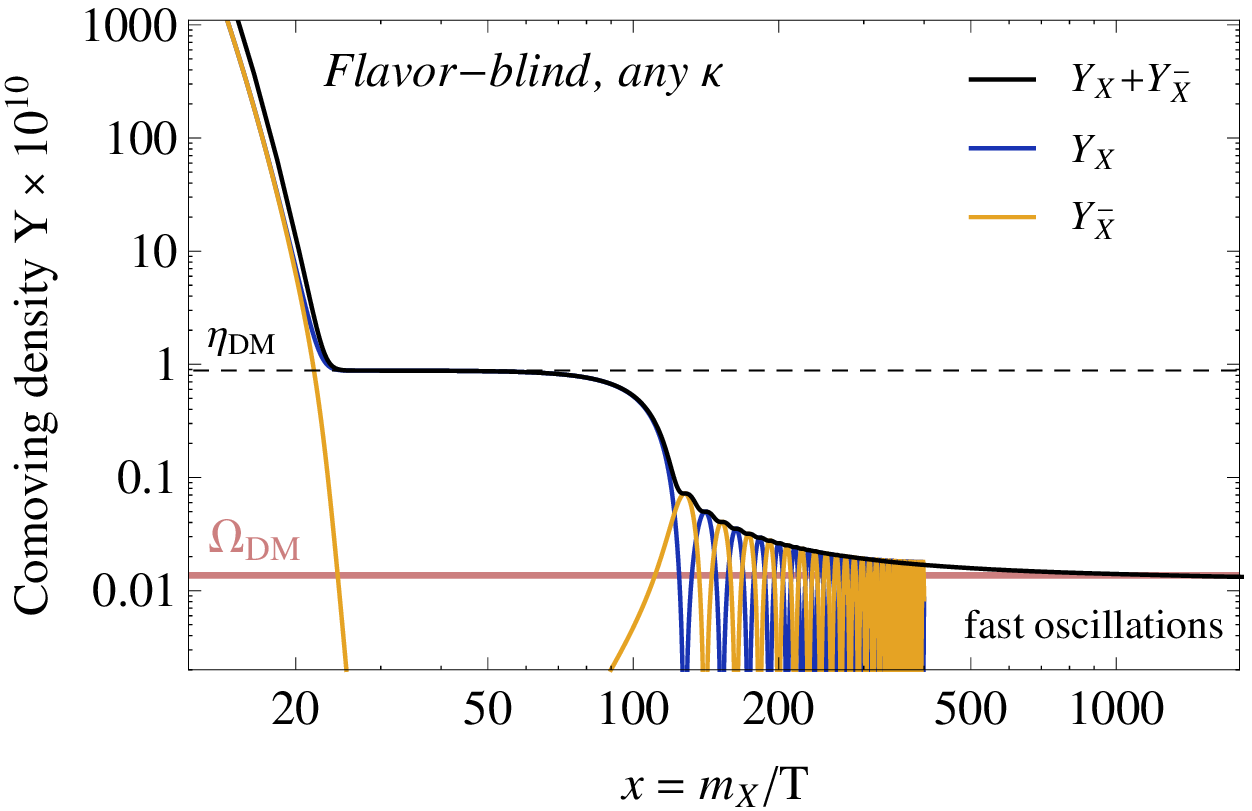} 
\includegraphics[scale=0.6]{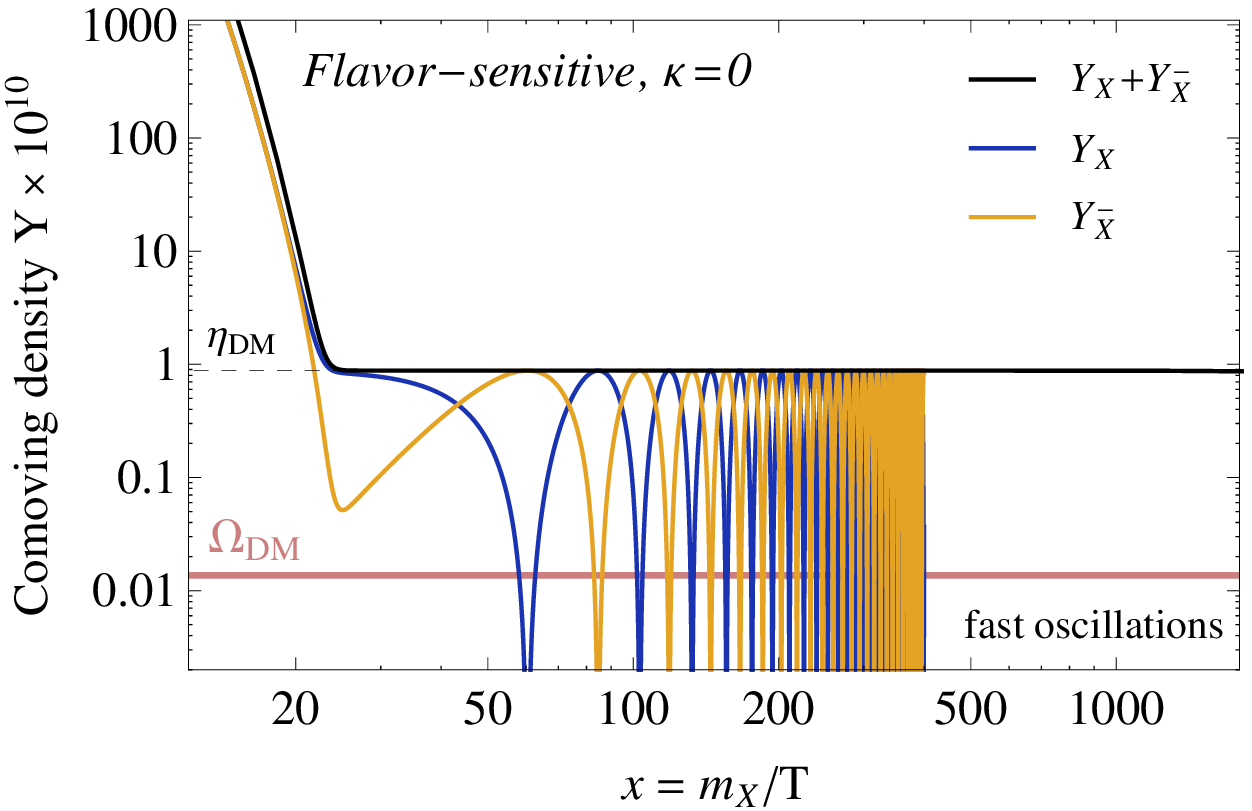} ~ \includegraphics[scale=0.6]{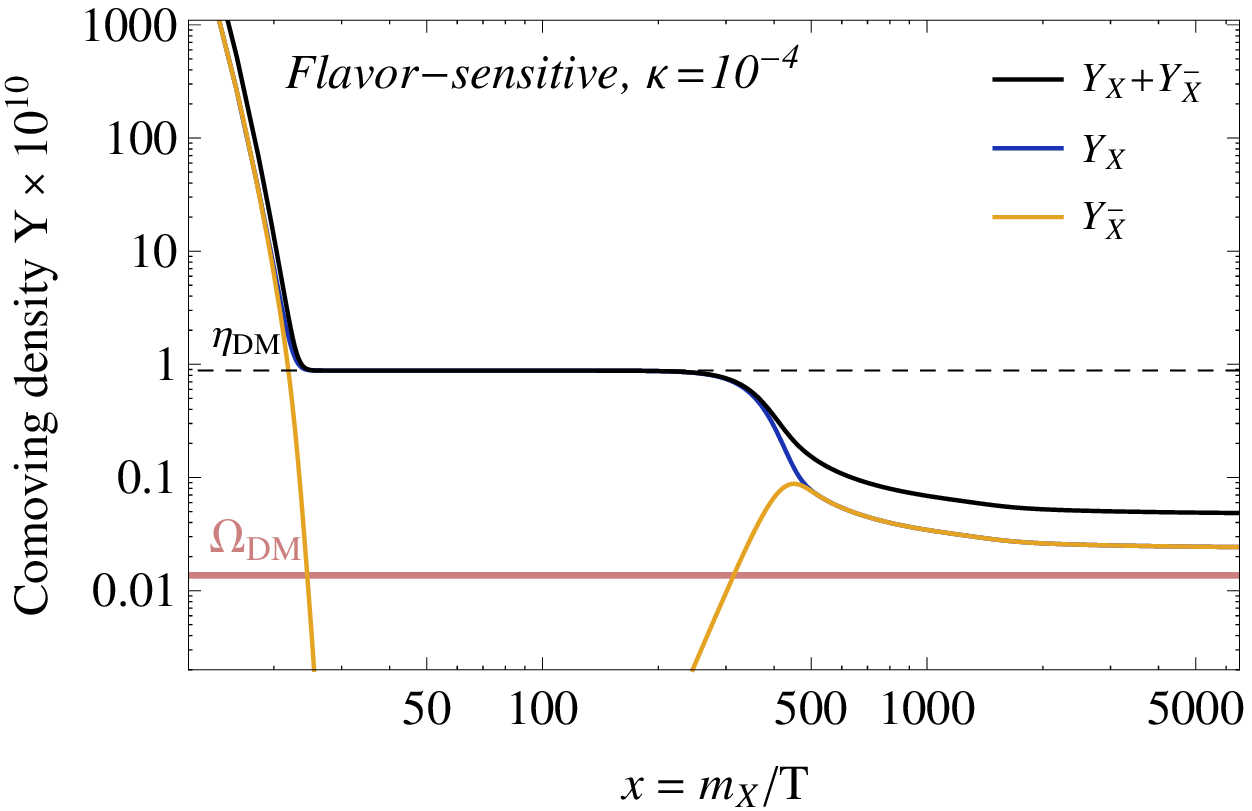} 
\caption{\it \label{plots1} Evolution of DM density for $m_X = 300$ GeV, $\langle \sigma v \rangle = 7.5$ pb, $\delta m = 10^{-7}$ eV. Top left: rates $H$, $\omega_{\rm osc}$, and $\Gamma_\pm$, for $\kappa = 10^{-4}$.  Top right: flavor-blind interaction for both $\kappa = 0$ (no scattering) and $\kappa = 10^{-4}$ (with scattering).  Bottom left: flavor-sensitive interaction with no scattering.  Bottom right: flavor-sensitive interaction with scattering.  Dashed line is initial DM asymmetry $\eta_{DM} = 8.8 \times 10^{-11}$.  Pink band is observed $\Omega_{DM}$.
}
\end{center}
\end{figure}

In Fig.~\ref{plots1}, we show the evolution of the DM density for an example case with $m_{X} = 300$ GeV, $\delta m = 10^{-7}$ eV, and $\sv = 7.5$ pb (assuming $s$-wave annihilation).
We set the scattering rate to be $\Gamma_\pm \equiv \kappa \,G_F^2 T^5$~\cite{Cirelli:2011ac}, where $G_F$ is the Fermi constant and $\kappa$ is a numerical coefficient.  
\begin{itemize}
\item {\it Top left:} Comparison of the Hubble rate $H$, oscillation rate $\omega_{\rm osc}$, and scattering rate $\Gamma_\pm$ for $\kappa = 10^{-4}$.  Asymmetric freeze-out occurs at $x \sim 20$, and without collisions, oscillations turn on when $\omega_{\rm osc} \sim H$, corresponding to $x \sim 30$.  With flavor-sensitive scattering, oscillations turn on when $\omega_{\rm osc} > \Gamma_-$ due to the quantum Zeno effect.  
\item {\it Top right:} Flavor-blind interaction case, with or without scattering.  Residual annihilation turns on when oscillations begin\footnote{Here, flavor-blind annihilation causes decoherence~\cite{Cirelli:2011ac}, delaying the onset of oscillations until $x\sim 100$.}, depleting the DM density by $\mathcal{O}(100)$.  A non-vanishing rate $\Gamma_+$ does not affect the DM evolution.
\item {\it Bottom left:} Flavor-sensitive interaction case, without scattering ($\kappa = 0$).  Oscillations turn on at $x \sim 30$, but no residual annihilation takes place.  The total DM density remains frozen-out at its asymmetric value.
\item {\it Bottom right:} Flavor-sensitive interaction case, with scattering ($\kappa = 10^{-4}$).  Scattering quenches oscillations until $x \sim 500$.  For $\omega_{\rm osc} > \Gamma_- > H$ ($x \gtrsim 500$), rapid oscillations and scatterings cause decoherence, and residual annihilation depletes the DM density by $\mathcal{O}(10)$.
\end{itemize}
The dashed line denotes the initial asymmetric DM charge density $\eta_{DM} \equiv Y_X - Y_{\bar X}$, assumed to be $\eta_{DM} = 8.8 \times 10^{-11}$, equal to the baryon density.  The pink band corresponds to the observed DM energy density $\Omega_{DM}$ (with $\pm2 \sigma$ thickness).

Residual annihilation is most efficient for a flavor-blind interaction, giving enough DM washout to reproduce the observed DM density for the parameters chosen here.  For a flavor-sensitive interaction with scattering, DM washout is reduced since the onset of oscillations is delayed (although significant washout is possible for larger $\sv$).  For a flavor-sensitive interaction with negligible scattering, this mechanism is inoperative, and $\Omega_{DM} = m_X \eta_{DM}s_0/\rho_c$ is fixed by the initial asymmetry, where $s_0$ and $\rho_c$ are the present entropy density and critical density respectively.  The latter two cases overproduce the DM density.

\begin{figure}[t]
\begin{center}
 \includegraphics[scale=0.6]{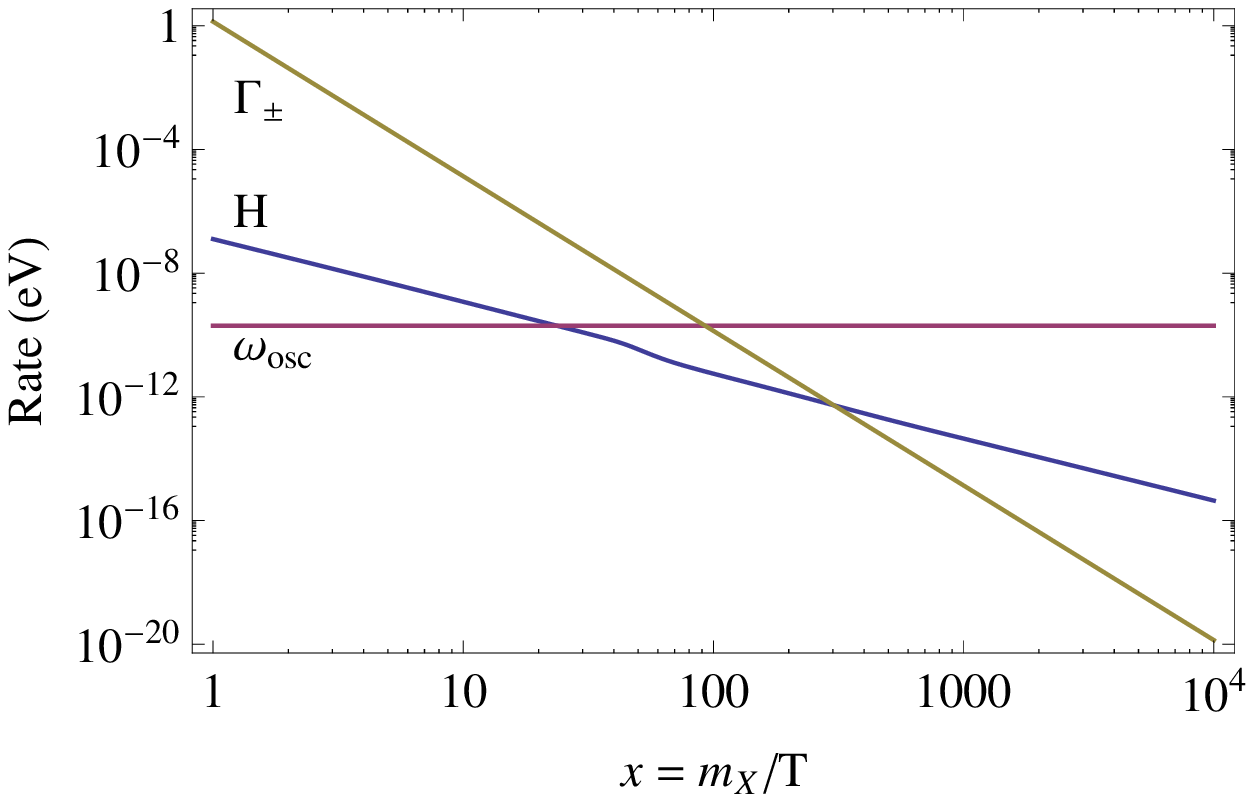} ~ \includegraphics[scale=0.6]{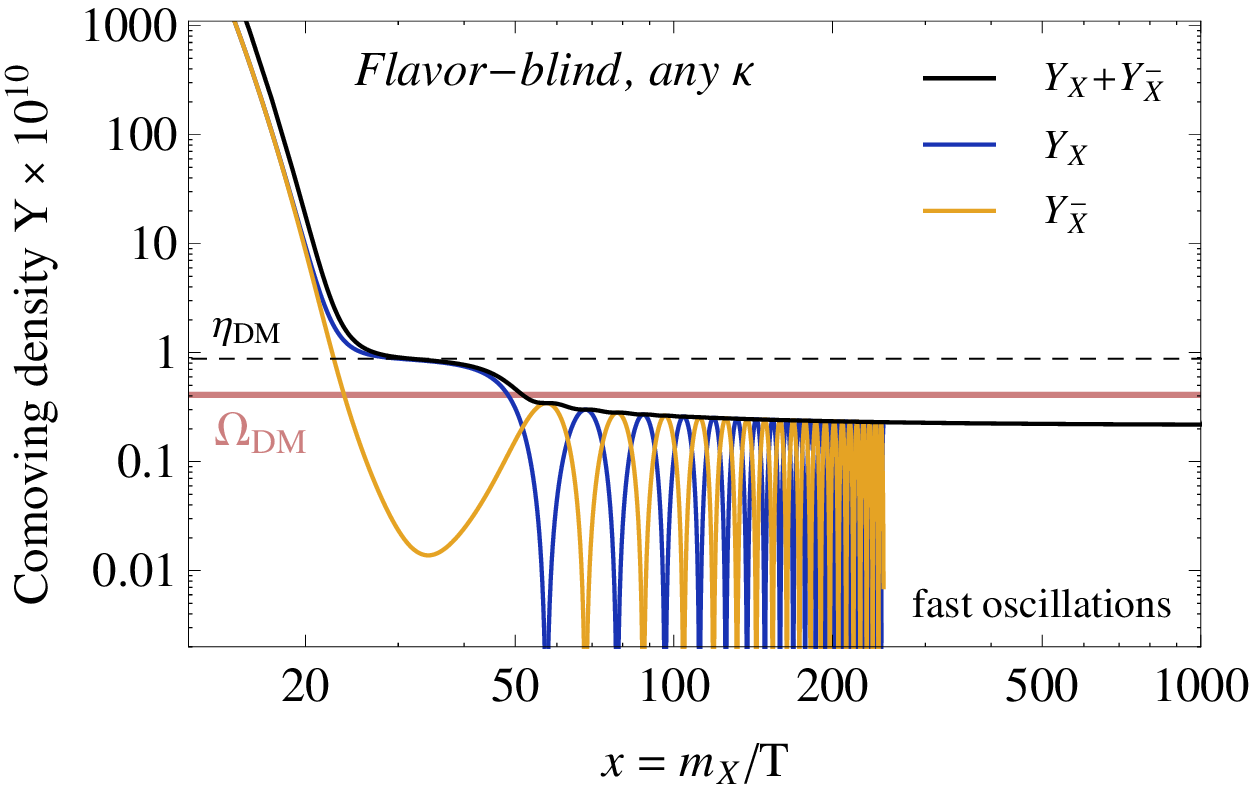}
 \includegraphics[scale=0.6]{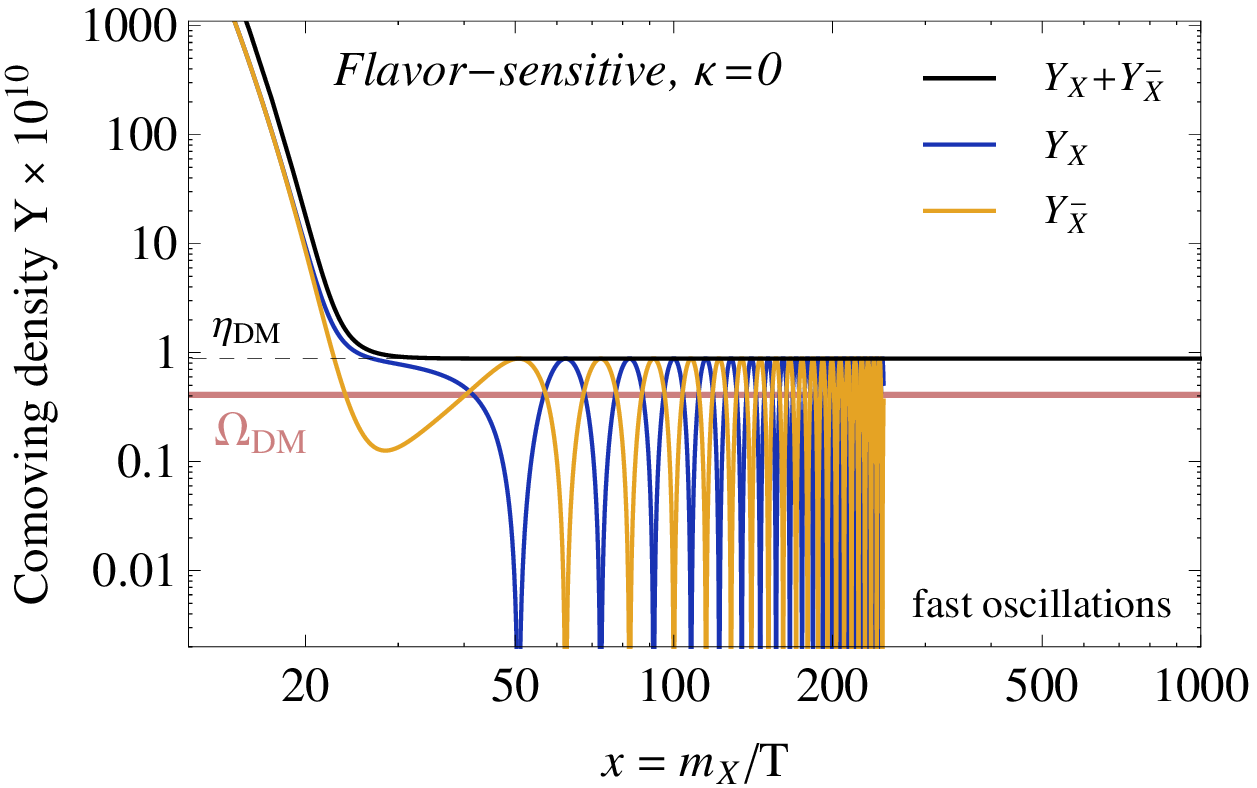} ~ \includegraphics[scale=0.6]{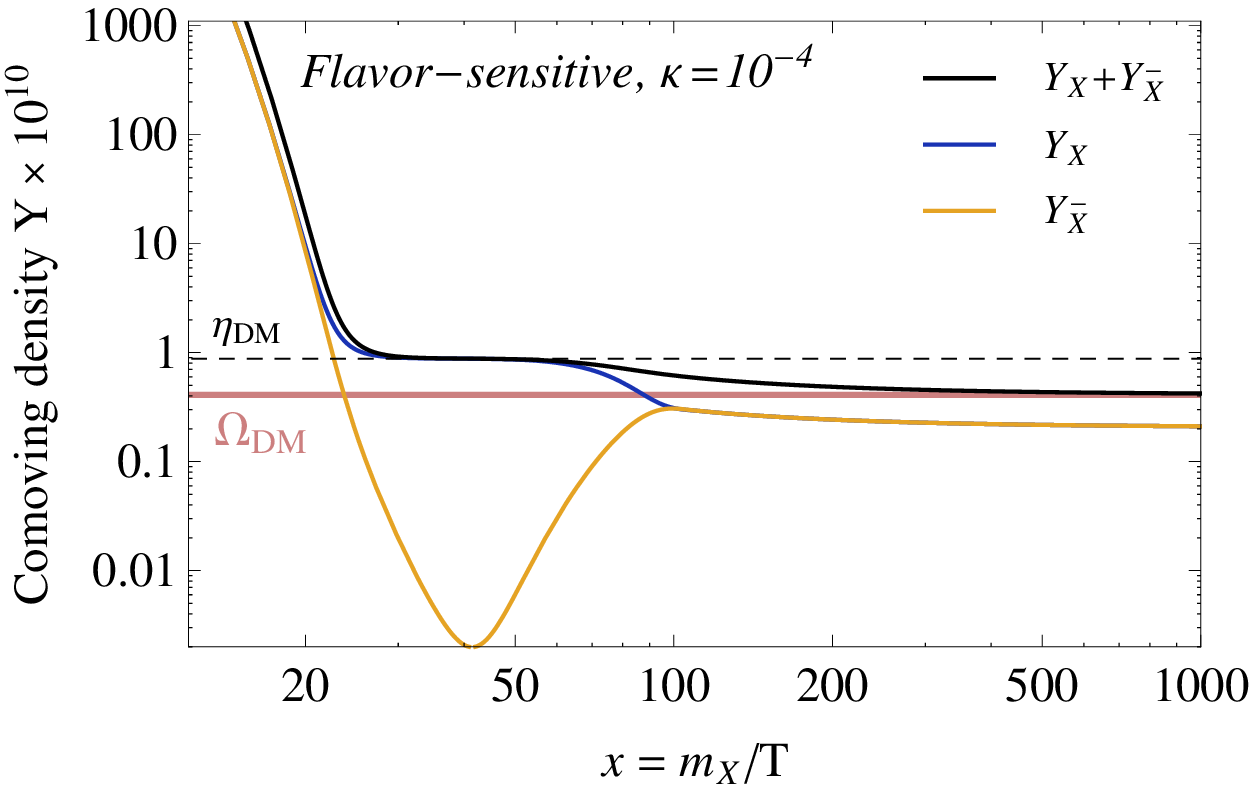}
\caption{\it \label{plots2} Evolution of DM density for $m_X = 10$ GeV, $\langle \sigma v \rangle = 5$ pb, $\delta m = 10^{-10}$ eV. Top left: rates $H$, $\omega_{\rm osc}$, and $\Gamma_\pm$, for $\kappa = 10^{-4}$.  Top right: flavor-blind interaction for both $\kappa = 0$ (no scattering) and $\kappa = 10^{-4}$ (with scattering).  Bottom left: flavor-sensitive interaction with no scattering.  Bottom right: flavor-sensitive interaction with scattering.  Dashed line is initial DM asymmetry $\eta_{DM} = 8.8 \times 10^{-11}$.  Pink band is observed $\Omega_{DM}$.}
\end{center}
\end{figure}

In Fig.~\ref{plots2}, we show the evolution of the DM density for another example with smaller DM mass: $m_{X} = 10$ GeV, $\delta m = 10^{-10}$ eV, and $\sv = 5$ pb (assuming $s$-wave annihilation).  The different panels correspond to the separate cases in Fig.~\ref{plots1}.  Since DM is lighter, less residual annihilation is required to reproduce the observed DM relic density, occuring here for the flavor-sensitive case with scattering ($\kappa = 10^{-4}$).  The flavor-blind case gives too much washout, favoring a heavier DM mass and/or smaller $\sv$, while the flavor-sensitive case with no scattering again gives $\Omega_{DM} = m_X \eta_{DM}s_0/\rho_c$.

Similar results were presented in Ref.~\cite{Cirelli:2011ac}.  We emphasize that for two cases --- flavor-blind annihilation without scattering ($O_+$, with $\kappa = 0$) and flavor-sensitive annihilation with scattering ($O_-$, with $\kappa \ne 0$) --- our results agree with theirs (despite differences in how the collision term couples to the components $Y_{ij}$).  For other cases, our results are qualitatively different and affect the relic DM density by an order of magnitude.

\subsection{Analytical Analysis for Flavor-Sensitive Annihilation}

It is possible to demonstrate analytically that flavor-sensitive annihilation is not reactivated by coherent oscillations, confirming our results above.  Following Ref.~\cite{Cirelli:2011ac}, we express the density matrix equation~\eqref{be3} as a system of coupled differential equations in terms of the variables
\be
\Sigma(x) \equiv Y_{11}+Y_{22} \,, \quad  \Delta(x) \equiv Y_{11}-Y_{22} \,,\quad  \Xi(x) \equiv Y_{12}-Y_{21} \, ,  \quad \Pi(x) \equiv Y_{12}+Y_{21} \; .
\ee
For the case of flavor-sensitive annihilation, with negligible scattering, Eq.~\eqref{be3} becomes 
\begin{align} \label{eq:be5}
\Sigma'=-\frac{2\sv }{H x}\left[\frac{1}{4}\left(\Sigma^2 -\Delta^2 -\Pi^2 +\Xi^2 \right)-Y^2_{\rm eq}\right]  , \;\;
\Delta'=\frac{2i\dm}{H x}\, \Xi \, , \;\;
\Xi'=\frac{2i\dm}{H x}\, \Delta\, ,\;\;  \Pi'=0 \, .
\end{align}
We take as the initial condition that the DM densities are frozen-out to their asymmetric values
$\Sigma(x_f) = \Delta(x_f) = \eta_{DM}$ at $x=x_f \sim 20$ (with $\Pi(x_f) = \Xi(x_f) = 0$).
Assuming $H \propto x^{-2}$, the equation for $\Delta(x)$ can be written as
\begin{eqnarray}
\Delta''=\frac{\Delta'}{x}-\frac{4\dm^2 \Delta}{H^2 x^2} \; ,
\label{eq:delta}
\end{eqnarray} 
which is satisfied for $\Delta(x)= \eta_{DM}  \cos( \delta m/H)$.
Through similar arguments, we also have $\Xi(x) = i \eta_{DM}  \sin( \delta m/H)$, and trivially $\Pi(x) = 0$.

From Eq.~\eqref{eq:be5}, the total DM density $\Sigma$ reaches its asymptotic solution when
\be
\Sigma^2=\Delta^2+\Pi^2-\Xi^2 \; ,
\ee
neglecting $Y_{\textrm{eq}}$ for $x \gg 1$.  However, plugging in our solutions, we find that the right-hand side is constant, given by $\Delta^2 + \Pi^2 - \Xi^2 = \eta_{DM}^2$.  Even in the presence of oscillations, the total density $\Sigma = Y_X + Y_{\bar{X}}$ remains frozen-out, fixed to its asymmetric freeze-out value --- even though the individual densities
\be
Y_X = \frac{\Sigma + \Delta}{2} = \eta_{DM} \cos^2 \left( \frac{\delta m}{2 H} \right) 
\; , \quad Y_{\bar X} = \frac{\Sigma - \Delta}{2} = \eta_{DM} \sin^2 \left( \frac{\delta m}{2 H} \right) 
\ee
do oscillate as expected.  (Setting $H=1/(2t)$, we obtain the standard oscillation formulae.)

\subsection{Indirect detection}

Although annihilation signals are typically quenched in ADM models, they can become reactivated in the presence of DM oscillations.  Since large annihilation cross sections are required to deplete the symmetric DM density, the resulting constraints can be important, but are highly dependent on the DM mass and final state channels.  A detailed analysis~\cite{Cirelli:2011ac} is beyond the scope of this work, and instead we briefly summarize some important points.

Studies of Big Bang nucleosynthesis (BBN) and the cosmic microwave background (CMB) constrain energy injection from DM during their respective epochs~\cite{Iocco:2008va}, $t_{\rm BBN}\sim 0.1~\textrm{sec}- 1$ min and $t_{\rm CMB} \sim 10^5$ yr.  Indirect detection signals from $\gamma$-ray and cosmic ray observations constrain DM annihilation during the present epoch, $t_{0} \sim 10^{10}$ yr~\cite{Cirelli:2010xx}.  For a flavor-blind interaction, annihilation commences when oscillations begin, for $t \gtrsim \delta m^{-1}$.  But for a flavor-sensitive interaction, annihilation occurs much later, when thermal effects cause decoherence, for $t \gtrsim \delta m^{-1}/v^{2}$.
In galactic systems (relevant for indirect detection), the typical velocity is $v \sim 10^{-3}$.  During the BBN and CMB epochs, we estimate the DM velocity as 
\be
v^2 \sim \frac{T^2}{T_{\rm kd} m_X} \sim \left\{ \begin{array}{cc} 
10^{-4}\times \left( \frac{T}{\rm MeV} \right)^2 \left( \frac{\rm GeV}{m_X} \right) \left(\frac{10 \, {\rm MeV}}{T_{\rm kd}}\right) &\; {\rm BBN} \\
10^{-16}\times \left( \frac{T}{\rm eV} \right)^2 \left( \frac{\rm GeV}{m_X} \right) \left(\frac{10 \, {\rm MeV}}{T_{\rm kd}}\right) &\; {\rm CMB} \end{array} \right.
\ee
where $T_{\rm kd}$ is the DM kinetic decoupling temperature.  

Ref.~\cite{Buckley:2011ye} infers strong indirect limits on oscillating ADM by requiring the oscillation time $\delta m^{-1}$ be larger than the time scales $t_{\rm BBN}, t_{\rm CMB}, t_0$ relevant for symmetric DM annihilation constraints.  (Clearly, these bounds are model-dependent.)  We emphasize that these constraints only apply for flavor-blind annihilation.  For flavor-sensitive annihilation, the bounds are weaker by $\sim 6 - 16$ orders of magnitude!

\section{Conclusions}
\label{sec:conc}

We have studied the impact of DM particle-antiparticle oscillations, generated by a DM number-violating Majorana-type mass, for asymmetric dark matter scenarios.  Oscillations erase the DM asymmetry, thereby reactivating annihilation after freeze-out, which 
can deplete the relic DM density and allow for indirect detection signals.  Such Majorana mass terms are a generic feature of any ADM model, greatly expand the ADM model-building possibilities, and provide a natural bridge between symmetric ({\it i.e.}, WIMP) and asymmetric DM.

Several previous works have considered DM oscillations, within specific models~ \cite{Cohen:2009fz,Cai:2009ia,Arina:2011cu,Falkowski:2011xh,Blum:2012nf} and in more general analyses~\cite{Buckley:2011ye,Cirelli:2011ac}.  Here, we provided the first rigorous derivation (from finite temperature field theory) of the density matrix equation of motion describing DM freeze-out, oscillations, and collisional processes.  We showed that oscillating DM exhibits particle-vs-antiparticle ``flavor'' effects, analogous to similar phenomena in the context of neutrino oscillations in a medium.  DM interactions can be ``flavor-blind'' or ``flavor-sensitive'' depending on how the interaction transforms under charge-conjugation of the DM field.  Flavor-sensitive interactions include DM scattering or annihilation through a new vector boson, while flavor-blind interactions include DM scattering or annihilation through a new scalar boson, or $t$-channel annihilation to two bosons.  
Our results agree with those in Ref.~\cite{Cirelli:2011ac} for the case of flavor-blind annihilation with flavor-sensitive or no scattering, but for other interactions these flavor effects lead to important qualitative differences.

The interplay of coherent oscillation and decoherence via scattering gives rise to a subtle combination of possible evolutions of ADM once the DM antiparticle state becomes populated.  The main new points emphasized in this paper are as follows:
\begin{itemize}
\item Once coherent oscillations commence, and the antiparticle becomes repopulated, DM annihilation only occurs via flavor-blind interactions, in the absence of coherence destroying scattering. 
\item Only flavor-sensitive scattering causes decoherence.  If scattering occurs only through flavor-blind interactions, scattering has no effect on ADM evolution.
\item If coherence is lost via flavor-sensitive scattering, flavor-sensitive annihilation may proceed.  (Flavor-blind annihilation occurs anytime after oscillations begin.)
\end{itemize}
We presented several arguments and numerical examples to demonstrate these conclusions.

There remains a rich phenomenology to explore in the presence of ADM oscillations, which we have only lightly touched on in this work.  In particular, over long times coherence can be lost through DM reheating during structure formation or late-time scattering.  What is clear is that while the ADM density may be fixed in the Universe by a DM asymmetry, there are a wide variety of scenarios to explore for the DM asymmetry at late times, leading in some cases to indirect detection signals for ADM.

\begin{acknowledgments}

We thank J.~Kearney, S.~Profumo, G.~Servant and T.~Volansky for helpful discussions.  ST would also like to thank V.~Cirigliano, C.~Lee, and M.~Ramsey-Musolf for collaboration on density matrix equations in the CTP formalism.  HBY and KZ are supported by NSF CAREER award PHY1049896 and by NASA Astrophysics Theory  grant NNX11AI17G.  ST is supported by DOE Grant \#DE-FG02-95ER40899.

\end{acknowledgments}

\appendix

\section{Nonequilibrium field theory derivation}

The closed-time-path (CTP), or real-time, nonequilibrium field theory formalism~\cite{Schwinger:1960qe} provides a useful and rigorous tool for deriving Boltzmann equations~\cite{Calzetta:1986cq}.  In this appendix, we use these methods to derive a Boltzmann-like equation for the density matrix describing DM freeze-out and oscillations, following Ref.~\cite{Cirigliano:2009yt,inprog}.  Similar methods have been adopted in other cosmological contexts~\cite{Riotto:1995hh,Garbrecht:2003mn,Prokopec:2003pj,Konstandin:2004gy,
Anisimov:2010dk,Beneke:2010dz,Beneke:2010wd}.

We consider the case where the DM field $X$ is a fermion, described by Eq.~\eqref{eq:L}.  (The arguments and results for the scalar DM case are similar.)   The basic building blocks are the thermally-averaged Green's functions
\be \label{GF}
S^<_{ij}(x,y)_{\alpha\beta} = - \left\langle  \bar{\Psi}_{j\beta}(y) \frac{}{}\Psi_{i\alpha}(x)  \right\rangle \: , \quad
S^>_{ij}(x,y)_{\alpha\beta} = \left\langle \Psi_{i\alpha}(x) \frac{}{}\bar{\Psi}_{j\beta}(y) \right\rangle 
\ee
where $\alpha,\beta$ are Dirac indices, and $x,y$ are spacetime coordinates (we assume flat spacetime for now).  ``Flavor'' indices $i,j$ label particle $\Psi_1 \equiv X$ and antiparticle $\Psi_2 \equiv X^C$.    Next, we define the average coordinate $\bar{x} \equiv \frac{1}{2}(x+y)$ and relative coordinate $r\equiv (x-y)$.  The Wigner transformation of $S^\gtrless(x,y)$ is given by
\be
S^\gtrless(k,\bar x) \equiv \int\! d^4 r \: e^{i \, k \cdot r} \,S^\gtrless(x,y) \; ,
\ee
which is simply a Fourier transform with respect to the relative coordinate $r$.

It turns out that $S^\gtrless(k,\bar{x})$ is closely related to the density matrix $\mathscr{F}_k$.  To see this connection, it is insightful to evaluate $S^\gtrless(k,\bar x)$ using a free-field mode expansion for $X$, setting $m_M = 0$:
\begin{subequations}
\begin{align}
X_{\alpha}(x) &= \int \!\! \frac{ d^3 k}{(2\pi)^3} \frac{1}{2\omega_{k}} \sum_s \left( u_\alpha(\mathbf k,s) \, a_{\mathbf k,s} \, e^{-i \, k \cdot x} + v_\alpha(\mathbf k,s) \, b^\dagger_{\mathbf k,s} \, e^{i \, k \cdot x} \right) \\
X^C_{\alpha}(x) &= \int \!\! \frac{ d^3 k}{(2\pi)^3} \frac{1}{2\omega_{k}} \sum_s \left( u_\alpha(\mathbf k,s) \, b_{\mathbf k,s} \, e^{-i \, k \cdot x} + v_\alpha(\mathbf k,s) \, a^\dagger_{\mathbf k,s} \, e^{i \, k \cdot x} \right) \; .
\end{align}
\end{subequations}
The density matrix $\mathscr{F}_k$ is defined by the expectation values of creation and annihilation operators, with the appropriate normalization factor:
\begin{subequations} \label{Fdef} \bea
\big\langle a^\dagger_{\mathbf k, s} \, a^{}_{\mathbf k^\prime, s^\prime} \big\rangle & =& 2 \omega_{k} \, \delta_{ss^\prime} \, (2\pi)^3 \delta^3(\mathbf k - \mathbf k^\prime) \, (\mathscr{F}_k)_{11} \\
\big\langle b^\dagger_{\mathbf k, s} \, b^{}_{\mathbf k^\prime, s^\prime} \big\rangle & =& 2 \omega_{k} \, \delta_{ss^\prime} \, (2\pi)^3 \delta^3(\mathbf k - \mathbf k^\prime) \, (\mathscr{F}_k)_{22} \\
\big\langle b^\dagger_{\mathbf k, s} \, a^{}_{\mathbf k^\prime, s^\prime} \big\rangle & =& 2 \omega_{k} \, \delta_{ss^\prime} \, (2\pi)^3 \delta^3(\mathbf k - \mathbf k^\prime) \, (\mathscr{F}_k)_{12} \\
\big\langle a^\dagger_{\mathbf k, s} \, b^{}_{\mathbf k^\prime, s^\prime} \big\rangle & =& 2 \omega_{k} \, \delta_{ss^\prime} \, (2\pi)^3 \delta^3(\mathbf k - \mathbf k^\prime) \, (\mathscr{F}_k)_{21}  \; .
\eea\end{subequations}
In Eq.~\eqref{Fdef}, we have assumed that the $X,X^C$ ensemble is rotationally invariant (depending only on $k \equiv |\mathbf k|$) and is uncorrelated with respect to spin (hence, $\delta_{ss^\prime}$).\footnote{More generally, $\mathscr{F}$ also carries spin indices $s,s^\prime$ and depends on the momentum direction $\hat{\mathbf k}$.  A more general derivation of fermionic density matrix equations using CTP methods is presented in Ref.~\cite{inprog}, allowing for spin coherence and momentum anisotropy in the density matrix.}  Plugging in everything, one finds
\begin{subequations}\label{Sgtrless}
\begin{align} 
S^<(k,\bar x) &= - (2\pi) \delta(k^2 - m_X^2)\, (\diracslash{k} + m_X) \left[ \theta(k^0) \, \mathscr{F}_k - \theta(-k^0)\, \left(\mathbbm{1}-\bar{\mathscr{F}}_k\right) \right] \\
S^>(k,\bar x) &= (2\pi) \delta(k^2 - m_X^2)\, (\diracslash{k} + m_X) \left[ \theta(k^0) \, \left(\mathbbm{1}-\mathscr{F}_k\right) - \theta(-k^0)\, \bar{\mathscr{F}}_k \right]
\end{align}\end{subequations}
where the form of $\bar{\mathscr{F}}_k$, given in Eq.~\eqref{barredF}, is fixed by relations in Eq.~\eqref{Fdef}.  

The starting point to obtain the equation of motion for $\mathscr{F}_k$ is the Schwinger-Dyson equations:
\begin{subequations} \label{SDeq}
\begin{align}
\widetilde S(x,y) &= \widetilde S^{(0)}(x,y) - i \int \!d^4 w \int \!d^4 z \; \widetilde S^{(0)}(x,w)\, \widetilde \Sigma(w,z) \, \widetilde S(z,y) \\
&= \widetilde S^{(0)}(x,y) - i \int\! d^4 w \int\! d^4 z \; \widetilde S(x,w) \, \widetilde \Sigma(w,z)\, \widetilde S^{(0)}(z,y) \; .
\end{align}
\end{subequations}
 In the CTP formalism, the fermionic Green's functions $\widetilde S$ and self-energies $\widetilde \Sigma$ (evaluated below) are expressed in matrix form as
\be
\widetilde S \equiv \left( \ba{cc} S^t & - S^< \\ S^> & - S^{\bar t} \ea \right) \; , \quad 
\widetilde \Sigma \equiv \left( \ba{cc} \Sigma^t & - \Sigma^< \\ \Sigma^> & - \Sigma^{\bar t} \ea \right) \; ,
\ee
where each component is a $4\times 4$ matrix in Dirac space and a $2\times 2$ matrix in flavor space.  The time-ordered $(\mathbbm{T})$ and anti-time-ordered $(\bar{\mathbbm{T}})$ Green's functions are
\begin{subequations}
\begin{align}
S^t_{ij}(x,y)_{\alpha\beta} &\equiv \left\langle \mathbbm{T}\frac{}{} \Psi_{i\alpha}(x) \frac{}{}\bar{\Psi}_{j\beta}(y) \right\rangle
= \theta(x^0 - y^0) \,S^>_{ij}(x,y)_{\alpha\beta} + \theta(y^0-x^0)\, S^<_{ij}(x,y)_{\alpha\beta}  \\
S^{\bar t}_{ij}(x,y)_{\alpha\beta} &\equiv \left\langle \bar{\mathbbm{T}}\frac{}{} \Psi_{i\alpha}(x) \frac{}{}\bar{\Psi}_{j\beta}(y) \right\rangle
= \theta(y^0 - x^0) \,S^>_{ij}(x,y)_{\alpha\beta} + \theta(x^0-y^0) \,S^<_{ij}(x,y)_{\alpha\beta} \; .
\end{align}
\end{subequations}
The free propagator $\widetilde{S}^{(0)}$, which we already computed in Eq.~\eqref{Sgtrless}, satisfies the free equations of motion
\begin{subequations}
\begin{align}
(i \overrightarrow{\diracslash{\partial}_x} - M)\,\widetilde  S^{(0)}(x,y) &= i \delta^4(x-y) \\
\widetilde S^{(0)}(x,y) \, (i \overleftarrow{\diracslash{\partial}_y} + M)  &= - i \delta^4(x-y)
\end{align}
\end{subequations}
The right-hand side is proportional to the identity in Dirac, flavor, and CTP propagator space.  If we act with the Dirac operator on the $\gtrless$-component of the Schwinger-Dyson equations \eqref{SDeq}, we obtain the Kadanoff-Baym equations:
\begin{subequations} \label{KBeq}
\begin{align}
(i \overrightarrow{\diracslash{\partial}_x} - M)\,S^\gtrless(x,y) &= \int \!d^4 z \; \left[ \widetilde \Sigma(x,z) \, \widetilde{S}(z,y) \right]^\gtrless \\
S^\gtrless(x,y) \, (i \overleftarrow{\diracslash{\partial}_y} + M)  &= - \int \!d^4 z \; \left[ \widetilde{S}(x,z) \, \widetilde \Sigma(z,y) \right]^\gtrless \; .
\end{align}
\end{subequations}
Taking the Wigner transformation of Eqs.~\eqref{KBeq}, we obtain
\begin{subequations} \label{kineq}
\begin{align}
\Big( \diracslash{k} - M + \frac{i}{2} \,\overrightarrow{\diracslash{\partial}_{\bar x}} \Big)  S^\gtrless &= e^{-i \Diamond}\Big(  \Sigma^h S^\gtrless + \Sigma^\gtrless S^h + \frac{1}{2} \Sigma^> S^< - \frac{1}{2} \Sigma^< S^> \Big) \\ 
S^\gtrless\Big( \diracslash{k} - M - \frac{i}{2} \, \overleftarrow{\diracslash{\partial}_{\bar x}} \Big)   &= e^{-i \Diamond}\Big( S^\gtrless  \Sigma^h + S^h \Sigma^\gtrless  + \frac{1}{2} S^< \Sigma^> - \frac{1}{2} S^> \Sigma^<  \Big) \; ,
\end{align} 
\end{subequations} 
where all $S$'s and $\Sigma$'s are Wigner-transformed functions of $(k,\bar x)$.  We also define $S^h \equiv S^t - S^{\bar{t}}$ and $\Sigma^h \equiv \Sigma^t - \Sigma^{\bar{t}}$.  The $\Diamond$ operator is defined by
\be
\Diamond\left( A(k,\bar x) B(k,\bar x) \right) \equiv \frac{1}{2} \left( \frac{\partial A}{\partial \bar x^\mu} \frac{\partial B}{\partial k_\mu} - \frac{\partial A}{\partial k_\mu} \frac{\partial B}{\partial \bar x^\mu} \right) \; ,
\ee
for two arbitrary Wigner-transformed functions $A$ and $B$.

Next, we simplify Eq.~\eqref{kineq} by making a number of assumptions.  First, we assume that quantities depend only on the time coordinate $t \equiv \bar{x}^0$, assuming spatial homogeneity and isotropy.  Second, we adopt a perturbative expansion in the self-energies $\Sigma$ and the oscillation parameter $\delta m$.\footnote{This scheme amounts to an expansion in the ratios of time scales, detailed in Ref.~\cite{Cirigliano:2009yt}.  The long time scales are: (i) the collisional mean-free-time $\tau_{\textrm{coll}}$, set by the interaction rate, and (ii) the oscillation time $\tau_{\textrm{osc}} \sim \delta m^{-1}$; the short time scale $\tau_{\textrm{int}}$ is corresponds to an ``intrinsic'' energy scale, set by $m_X^{-1}$ or $T^{-1}$.  We work in the regime $\tau_{\textrm{coll}},\tau_{\textrm{osc}} \gg \tau_{\textrm{int}}$, counting each power of $\Sigma$ as $\mathcal{O}(\tau_{\textrm{int}}/\tau_{\textrm{coll}})$ and each power of $\delta m$ as $\mathcal{O}(\tau_{\textrm{int}}/\tau_{\textrm{osc}})$.}

Working at zeroth order in $\Sigma$ and $\delta m$, Eq.~\eqref{kineq} becomes
\be \label{zeroth1}
\Big( \diracslash{k} - m_X + \frac{i \gamma^0}{2} \,\overrightarrow{\partial_{t}} \Big)  S^\gtrless(k,t)  =
S^\gtrless(k,t) \Big( \diracslash{k} - m_X - \frac{i \gamma^0}{2} \, \overleftarrow{\partial_{t}} \Big)  = 0 \; .
\ee
From Eq.~\eqref{zeroth1}, it is straight-forward to show that
\begin{align}
&\Big( \diracslash{k} - m_X + \frac{i \gamma^0}{2} \,\overrightarrow{{\partial}_{t}} \Big)^2  S^\gtrless(k,t)  =
\Big(k^2 - m^2 - \frac{1}{4} \, \partial_t^2 + i k^0 \partial_t \Big) S^\gtrless(k,t) = 0 \\
&S^\gtrless(k,t) \Big( \diracslash{k} - m_X - \frac{i\gamma^0}{2} \, \overleftarrow{{\partial}_{t}} \Big)^2 =  
\Big(k^2 - m^2 - \frac{1}{4} \, \partial_t^2 - i k^0 \partial_t \Big) S^\gtrless(k,t) = 0 \; .
\end{align}
Taking the sum and difference, we have
\be \label{zeroth2}
2 k^0 \partial_t S^\gtrless(k,t) = 0 \; , \quad \Big( k^2 - m_X^2 - \frac{1}{4} \partial_t^2 \Big) S^\gtrless(k,t) = 0 \; .
\ee
Eq.~\eqref{zeroth2} implies (for $k^0 \ne 0$) that $\partial_t S^\gtrless$ can be counted as first order in $\Sigma$ or $\delta m$, since it vanishes at zeroth order.\footnote{$k^0 = 0$ solutions correspond to coherent particle-antiparticle production~\cite{Herranen:2008hi}.  We neglect these modes in our analysis.}  Dropping the second order $\partial_t^2/4$ term, we see that $S^\gtrless$ vanishes unless $k^2 = m_X^2$.  Moreover, dropping the ${{\partial}}_{t}$ terms from Eq.~\eqref{zeroth1}, we have
\be \label{zeroth3}
(\diracslash{k} - m_X) S^\gtrless(k,t)  = S^\gtrless(k,t) ( \diracslash{k} - m_X )=0 \;  ,
\ee
which is satisfied if $S^\gtrless(k,t)$ is proportional to $(\diracslash{k} + m_X)$.\footnote{A more general Dirac structure is allowed by Eq.~\eqref{zeroth3} which parametrizes spin asymmetries~\cite{Prokopec:2003pj,Konstandin:2004gy,Anisimov:2010dk,Beneke:2010wd,Beneke:2010dz,Herranen:2008hi,Garbrecht:2011aw,Garny:2011hg}
or spin coherence in the density matrix~\cite{inprog}.  Here, we assume DM spins are unpolarized in the early Universe.}  We can implement these constraints explicitly by parametrizing $S^\gtrless$ as
\be \label{zeroth4}
S^\gtrless(k,t) = (2\pi) \delta(k^2 - m^2_X) \, (\diracslash{k} + m_X) \, \big[ \theta(k^0) \, g^\gtrless_+(k,t) + \theta(-k^0) \,  g^\gtrless_-(k,t) \big] \; .
\ee
The four unknown functions $g^\gtrless_\pm$ are not all independent.  The canonical anticommutation relations $\{ X_\alpha(t,\mathbf x), X^\dagger_\beta(t,\mathbf y) \} = \delta^3(\mathbf x - \mathbf y) \delta_{\alpha\beta}$ imply
\be
\frac{1}{4} \int^\infty_{-\infty} \frac{d k^0}{2\pi} \, \textrm{Tr}\left[ \gamma^0\left( S^>(k,t) - S^<(k,t)\right) \right] = \mathbbm{1}\, , \quad
\int^\infty_{-\infty} \frac{d k^0}{2\pi} \, \textrm{Tr}\left[  S^>(k,t) - S^<(k,t) \right] = 0 \, ,
\ee
where ``\textrm{Tr}'' traces over Dirac indices only (not flavor indices).  Plugging in Eq.~\eqref{zeroth4}, we have 
\be
g_+^>(k,t) - g_+^<(k,t) = g_-^>(k,t) - g_-^<(k,t) = \mathbbm{1} \; . \label{rels}
\ee
Defining $\mathscr{F}_k \equiv - g_+^<$ and $\bar{\mathscr{F}}_k \equiv g_-^>$, and using Eq.~\eqref{rels}, we reproduce our previous expression for $S^\gtrless(k,t)$ given in Eq.~\eqref{Sgtrless}.  Furthermore, $\mathscr{F}_k$ and $\bar{\mathscr{F}}_k$ are related by charge conjugation.  From the Green's functions' definitions in Eq.~\eqref{GF}, we have
\begin{subequations}
\begin{align}
S^>_{11}(k,t) &= - \mathcal{C} S^<_{22}(-k,t)^T \mathcal{C} \, , &
S^>_{22}(k,t) &= - \mathcal{C} S^<_{11}(-k,t)^T \mathcal{C} \, \\
S^>_{12}(k,t) &= - \mathcal{C} S^<_{12}(-k,t)^T \mathcal{C} \, , &
S^>_{21}(k,t) &= - \mathcal{C} S^<_{21}(-k,t)^T \mathcal{C} \,  \; ,
\end{align}
\end{subequations}
where $\mathcal{C} \equiv i \gamma^2 \gamma^0$.
Taking the solution for $S^\gtrless(k,t)$, we find that the form of $\bar{\mathscr{F}}_k$ is fixed according to Eq.~\eqref{barredF}.    

Dynamical evolution of the density matrix occurs at first order in $\delta m$ and $\Sigma$.  Taking the difference of Eqs.~\eqref{kineq}, and multiplying by $-i$, we have\footnote{The term $- i \big[ \Sigma^\gtrless, \, S^h \big]$ is $\mathcal{O}(\Sigma \times \delta m)$ and can be neglected~\cite{Cirigliano:2009yt,inprog}.}
\begin{align} \label{first}
\frac{1}{2} \left\{ \gamma^0 , \partial_{t} S^\gtrless \right\} =  i \left[ \diracslash{k} - M - \Sigma^h, \,  S^\gtrless \right]
   + \frac{i}{2} \big\{ \Sigma^<, \, S^> \big\}  - \frac{i}{2} \big\{ \Sigma^>, \, S^< \big\}  \; .
\end{align}
The evolution equation for the density matrix, given by Eq.~\eqref{be}, is obtained by taking the following ``moment'' of Eq.~\eqref{first}:
\be \label{moment}
- \int_0^\infty \frac{dk^0}{2\pi} \; \textrm{Tr}\left[ \left(\frac{\diracslash{k}+m_X}{4 m_X}\right) \textrm{Eq.}~\eqref{first} \; \right] \; .
\ee
The left hand side gives
\be \label{lhs}
- \int_0^\infty \frac{dk^0}{2\pi} \; \textrm{Tr}\left[ \left(\frac{\diracslash{k}+m_X}{4 m_X}\right) \frac{1}{2} \left\{ \gamma^0, \partial_t S^\gtrless(k,t) \right\} \right] = \frac{\partial \mathscr{F}_k}{\partial t} \; ,
\ee
where we have substituted in for $S^\gtrless(k,t)$ the zeroth order solution.  In general, $S^\gtrless$ can receive first order corrections to the form given in Eq.~\eqref{Sgtrless}, leading to a modification of the spectral function $\delta(k^2 - m_X^2)$ or possibly additional terms involving Dirac structures besides $(\diracslash{k}+m_X)$.  The latter do not contribute to Eq.~\eqref{lhs} since $\textrm{Tr}[(\diracslash{k}+m_X) ...]$ projects out only $(\diracslash{k}+m_X)$ terms.  Modifications to the spectral function can be neglected, since $\partial_t \mathscr{F}_k$ is already first order, and we truncate at this order.  Similarly, on the right hand side, we have
\be \label{rhs}
- \int_0^\infty \frac{dk^0}{2\pi} \; \textrm{Tr}\left[ \left(\frac{\diracslash{k}+m_X}{4 m_X}\right) i \left[ \diracslash{k} - M , S^\gtrless(k,t) \right] \right] = - i \left[ \mathcal{H}_k , \mathscr{F}_k \right] \; ,
\ee
with $\mathcal{H}_k$ defined in Eq.~\eqref{ham}.  Since Eq.~\eqref{rhs} is explicitly $\mathcal{O}(\delta m)$, we again use the zeroth order solution for $S^\gtrless$.  

\begin{figure}[t]
\begin{center}
\includegraphics{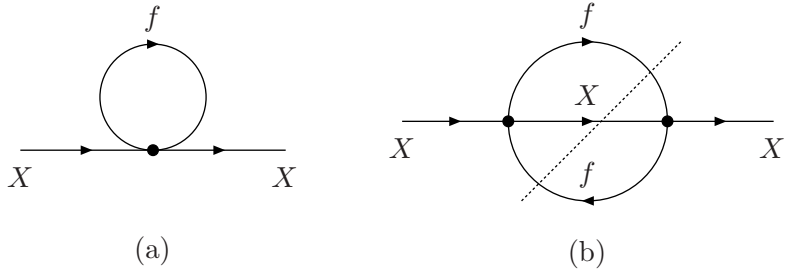}
\caption{\it Self-energy diagrams:
(a) Leading contribution to $\Sigma^h$ corresponds to a medium-induced mass term. (b) Leading contribution to $\Sigma^\gtrless$ at two-loop generates collision terms from tree-level scattering $X f \to X f$ and annihilation $X \bar{X} \to f \bar{f}$, with cut shown by the dotted line.\label{feyn}}
\end{center}
\end{figure}

Next, we evaluate $\mathcal{O}(\Sigma)$ collision terms appearing in Eq.~\eqref{first}.  We consider as an example a four-fermion contact interaction, with $\Lint$ given in Eq.~\eqref{contact}.  The self-energies can be computed perturbatively in $G_X$, with the leading contributions shown in Fig.~\ref{feyn}.  The $\Sigma^h$ term, arising at $\mathcal{O}(G_X)$, corresponds to the usual medium-induced shift in the mass matrix $M$, analogous to the MSW effect in neutrinos.  For nonrelativistic DM, this effect may be neglected.  The remaining terms correspond to $2 \to 2$ processes.  The leading $\mathcal{O}(G_X^2)$ contribution to $\Sigma^\gtrless$ is given by
\begin{align}
\Sigma^\gtrless(k,\bar x) = \frac{i}{2}\, {G^2_X} \int \! \frac{d^4 k^\prime}{(2\pi)^4} \int \! \frac{d^4 p}{(2\pi)^4}& \int \! \frac{d^4 p^\prime}{(2\pi)^4}
\: (2\pi)^4 \delta^4(k+p-k^\prime-p^\prime) \\
& \times O_\pm \,\Gamma^a \, S^\gtrless(k^\prime,\bar x)\, O_\pm \,\Gamma^b \:
\textrm{Tr}\left[S_f^\lessgtr(p) \, \Gamma_a \,S^\gtrless_f(p^\prime)\, \Gamma_b \right] \notag \; .
\end{align}
where $S_f^\gtrless(p)$ denotes the Green's functions for fermion $f$
\begin{subequations} \label{Sfgtrless}
\begin{align} 
S^>_f(p) &= (2\pi) \delta(p^2- m_f^2)\, (\diracslashb{p}+m_f) \big[ \theta(p^0) \, (1-f_p) - \theta(-p^0)\, \bar f_p \big] \\
S^<_f(p) &= - (2\pi) \delta(p^2-m_f^2)\, (\diracslashb{p}+m_f) \big[ \theta(p^0) \, f_p - \theta(-p^0)\, (1- \bar{f}_p) \big] \; .
\end{align}
\end{subequations}
Substituting these expressions into Eq.~\eqref{first}, it is straightforward to show that
\be 
- \int_0^\infty \frac{dk^0}{2\pi} \; \textrm{Tr}\left[ \left(\frac{\diracslash{k}+m_X}{4 m_X}\right) 
\frac{i}{2} \left( \big\{ \Sigma^<, \, S^> \big\}  -  \big\{ \Sigma^>, \, S^< \big\} \right)  \right] = \mathscr{C}_k[\mathscr{F}] ,
\ee
with collision term $\mathscr{C}_k$ given in Eqs.~\eqref{Cann} and \eqref{Cscat}.

In summary, Eq.~\eqref{moment} has become
\be \label{final}
\frac{\partial \mathscr{F}_k}{\partial t} = - i [ \mathcal{H}_k,\mathscr{F}_k] + \mathscr{C}_k[\mathscr{F}] \; .
\ee
Thus far, we have assumed flat spacetime.  In an expanding Friedmann-Robertson-Walker (FRW) spacetime, our results remain valid provided we replace physical time $t$ with conformal time $\eta$ (defined by $dt \equiv a \, d\eta$, where $a$ is the scale factor) and physical momentum $k$ with comoving momentum $k_{\textrm{co}} \equiv k a$, and we rescale dimensionful parameters by $a$ (e.g., $M \to a M$), as required by a canonically normalized kinetic term~\cite{Beneke:2010wd}.  Re-expressing the density matrix equation in terms of physical variables, the left hand side becomes
\be
\frac{1}{a} \frac{\partial \mathscr{F}_k}{\partial \eta} = \frac{\partial \mathscr{F}_k}{\partial t} + \frac{\partial \mathscr{F}_k}{\partial k} \frac{\partial k}{\partial t} = 
\frac{\partial \mathscr{F}_k}{\partial t} - H k \frac{\partial \mathscr{F}_k}{\partial k} \; ,
\ee
with Hubble constant $H$ (the right hand side is unchanged).  Incorporating an FRW spacetime thereby amounts to the replacement $\partial_t \mathscr{F}_k \to \partial_t \mathscr{F}_k - H k \,\partial_k \mathscr{F}_k$ compared to our flat spacetime results.  
Thus, we obtain the density matrix equation of motion given in Eq.~\eqref{be}.

\end{document}